\documentclass[prb, preprint, unsortedaddress, showpacs, nofootinbib, eqsecnum]{revtex4}

\usepackage{amsfonts, amsbsy, amssymb, amsmath, graphicx, color,  float}

\restylefloat{figure}
\newcommand{\rmd}{{\rm d}}

\newcommand{\calD}{\mathcal{D}}

\newcommand{\calM}{\mathcal{M}}

\newcommand{\qb}{\bar{q}}
\newcommand{\pb}{\bar{p}}
\newcommand{\bqb}{\bar{\boldsymbol{q}}}
\newcommand{\bpb}{\bar{\boldsymbol{p}}}

\newcommand{\Qb}{\bar{Q}}
\newcommand{\Pb}{\bar{P}}

\newcommand{\bv}{\boldsymbol{v}}

\begin{document}


\title{Index $k$ saddles and dividing surfaces in phase space, with applications to isomerization dynamics}



\author{Peter Collins}
\affiliation{School of Mathematics  \\
University of Bristol\\Bristol BS8 1TW\\United Kingdom}

\author{Gregory S. Ezra}
\email[]{gse1@cornell.edu}
\affiliation{Department of Chemistry and Chemical Biology\\
Baker Laboratory\\
Cornell University\\
Ithaca, NY 14853\\USA}

\author{Stephen Wiggins}
\email[]{stephen.wiggins@mac.com}
\affiliation{School of Mathematics and\\
Bristol Centre for Nanoscience and Quantum Information \\
University of Bristol\\Bristol BS8 1TW\\United Kingdom}


\date{\today}

\begin{abstract}
In this paper we continue our studies of the phase space  geometry and dynamics associated 
with index $k$ saddles ($k > 1$) of the potential energy surface.  Using Poincar\'e-Birkhoff 
normal form theory, we give an explicit formula for a ``dividing surface'' in phase space, 
i.e. a co-dimension one surface (within the energy shell) through which all 
trajectories  that ``cross'' the region of the index $k$ saddle
must pass.  With a generic non-resonance assumption, the normal form provides 
$k$ (approximate) integrals that describe the saddle dynamics in a neighborhood 
of the index $k$ saddle. These integrals provide a symbolic description 
of all trajectories that pass through a neighborhood of the saddle. 
We give a parametrization of the dividing surface which is used as the basis for a  
numerical method to sample the dividing surface.  Our 
techniques are applied to isomerization dynamics on  
a  potential energy surface  having  4 minima;
two symmetry related pairs of minima are connected by low energy index one saddles, with 
the pairs themselves connected via higher energy index one saddles and an 
index two saddle at the origin.
We compute and sample the dividing surface and show that our approach enables us to distinguish between 
concerted crossing (``hilltop crossing'') isomerizing trajectories
and those trajectories that are not concerted crossing (potentially sequentially isomerizing
trajectories).  We then consider the effect of additional ``bath modes''
on the dynamics, which is a four degree-of-freedom system.  For this system we 
show that the  normal form and dividing surface can be realized and sampled and that, 
using the approximate integrals of motion and our symbolic description of trajectories, 
we are able to choose initial conditions corresponding to concerted crossing isomerizing 
trajectories and (potentially) sequentially isomerizing trajectories. 
\end{abstract}


\maketitle


\section{Introduction}
\label{sec:intro}

Transition state theory has long been, and continues to be, a cornerstone
of the theory of chemical reaction rates \cite{Wigner38,Keck67,Pechukas81,Truhlar83,Anderson95,Truhlar96}.
A large body of recent research has shown that index one saddles \cite{saddle_footnote1}
of the potential energy surface \cite{Mezey87,Wales03}
give rise to a variety of geometrical structures in {\em phase space},
enabling the realization of Wigner's vision of a transition state theory
constructed in \emph{phase space}
\cite{Wiggins90,wwju,ujpyw,WaalkensBurbanksWiggins04,WaalkensWiggins04,WaalkensBurbanksWigginsb04,
WaalkensBurbanksWiggins05,WaalkensBurbanksWiggins05c,SchubertWaalkensWiggins06,WaalkensSchubertWiggins08,
MacKay90,Komatsuzaki00,Komatsuzaki02,Wiesenfeld03,Wiesenfeld04,Wiesenfeld04a,Komatsuzaki05,Jaffe05,Wiesenfeld05,Gabern05,Gabern06,Shojiguchi08}.

Following these studies, attention has naturally focussed on
phase space structures associated with saddles of index
greater than one, and their possible dynamical significance \cite{Ezra09,Haller10,Haller10a}.
In previous work we have described the phase space structures and their influence on transport
in phase space  associated with  {\em index two saddles} of the potential energy surface
for $n$ degree-of-freedom (DoF) deterministic, time-independent Hamiltonian systems \cite{Ezra09}.
(The case of higher index saddles has also been investigated by Haller et al. \cite{Haller10,Haller10a};
see also refs \onlinecite{Toda05,Toda08,Teramoto11}.)

The phase space manifestation of an index one saddle of the potential energy surface in an $n$ DoF
system is an equilibrium point of the associated Hamilton's
equations of saddle-center-$\ldots$-center stability type.
This means that the matrix associated with the linearization of Hamilton's equations about
the equilibrium point has one pair of real eigenvalues of equal magnitude, but opposite
in sign ($\pm \lambda$) and $n-1$ pairs of purely imaginary eigenvalues, $\pm i \omega_j$, $j=2, \ldots , n$.

The phase space manifestation of an index $k$ saddle is an
equilibrium point of saddle stability type:
\begin{equation}
\label{eq:saddle_center}
\underbrace{\text{saddle}\times\ldots\times\text{saddle}}_{k \;\; \text{times}}
\times \underbrace{\text{center} \times \ldots \times\text{center}}_{n-k \;\; \text{times}}.
\end{equation}
The matrix associated with the linearization of Hamilton's equations about
the equilibrium point then has $k$ pairs of real eigenvalues of equal magnitude, but opposite
in sign ($\pm \lambda_i$, $i=1,\ldots, k$) and $n-k$ pairs of purely imaginary eigenvalues,
$\pm i \omega_j$, $j=k+1, \ldots , n$ \cite{saddle_footnote2}.
Informally, an index $k=2$
saddle on a potential surface corresponds to a maximum or ``hilltop'' in the potential
\cite{Heidrich86,Mezey87,Wales03}.

Although it has been argued
on the basis of the Murrell-Laidler theorem \cite{Murrell68,Wales03}
that critical points of index $2$ and higher are of no
direct chemical significance \cite{Mezey87,Minyaev91},
many instances of index 2 (and higher) saddles of chemical
significance have been identified
\cite{Heidrich86}.
For example, Heidrich and Quapp discuss the case of
face protonated aromatic compounds, where high energy saddle points of index two prevent
proton transfer across the aromatic ring, so that proton shifts must occur at the ring periphery
\cite{Heidrich86}.  Index two saddles are found on potential surfaces
located between pairs of minima and index one saddles,
as in the case of internal rotation/inversion in the H$_2$BNH$_2$ molecule
\cite{Minyaev97} or in urea \cite{Bryantsev05},
or connected to index one saddles connecting
four symmetry related minima, as for  isomerization pathways in B$_2$CH$_4$ \cite{Fau95}.

Saddles with index $>1$ might well play a significant role in determining system properties
and dynamics
for low enough potential barriers \cite{Carpenter04,Bachrach07} or at high enough energies
\cite{Meroueh02}.
The role of higher index saddles in determining the behavior of supercooled liquids and glasses
\cite{Cavagna01,Cavagna01a,Doye02,Wales03a,Angelini03,Shell04,Grigera06,Coslovich07,Angelini08}
is a topic of continued interest, as is
the general relation between configuration space topology (distribution of saddles)
and phase transitions \cite{Kastner08}.

Several examples of non-MEP (minimum energy path) reactions
\cite{Mann02,Sun02,Debbert02,Ammal03,Carpenter04,Lopez07,Lourderaj08} and ``roaming'' mechanisms
\cite{Townsend04,Bowman06,Shepler07,Shepler08,Suits08,Heazlewood08}
have been identified in recent years; the dynamics of these reactions is not mediated by
a single conventional transition state associated with an index one saddle.
Higher index saddles can also become mechanistically important for structural transformations
of atomic clusters \cite{Ball96} when the range of the pairwise potential is reduced \cite{Berry96}.
We note for example the work of Shida on the importance of high index saddles
in the isomerization dynamics of Ar$_7$ clusters \cite{Shida05}.

The role of index two saddles in the (classical) ionization dynamics of the Helium atom in
an external electric field \cite{Eckhardt01,Sacha01,Eckhardt06} has
recently been studied from a phase space perspective by Haller et al.\ \cite{Haller10,Haller10a}.

In previous work we have discussed
phase space structures and their influence on phase space transport
in some detail for the case of an index two saddle
of the potential energy surface
corresponding to an equilibrium point
of saddle--saddle--center-$\ldots$--center stability type \cite{Ezra09}.
In this paper we extend our analysis in several respects.

One motivation for the work presented here is the possibility of
developing a dynamically based  characterization of
\emph{concerted} and \emph{sequential} reaction pathways \cite{Carpenter04,Bachrach07}
in phase space.
Consider isomerization dynamics on the
model potential shown in Fig.\ \ref{fig:seq_con}.
(This potential is discussed in more detail in Sec.\ \ref{sec:model_potential}.)
The potential has 4 minima;
two symmetry related pairs of minima are connected by low energy index one saddles, while
the pairs themselves are connected via higher energy index one saddles.
An index two saddle (hilltop, denoted $\ddagger\ddagger$) is located at the origin.
In Figure \ref{fig:seq_con},
we indicate schematically two possible pathways
between the lower left hand well (designated $(--)$; the symbolic code is
discussed further in Sec.\ \ref{sec:crossing}) and the upper right hand well, $(++)$ 
(see also Fig.\ 7 of ref.\ \onlinecite{Nguyen10}).
At energies above that corresponding to the index two saddle, there are,
qualitatively speaking, two possible isomerization routes: a \emph{sequential} path,
shown in Fig.\ \ref{fig:seq_con}a, in which a trajectory passes through
an intermediate well (either $(+-)$ or $(-+)$) in the course of the isomerization,
and a \emph{concerted} route, shown in Fig.\ \ref{fig:seq_con}b,
in which the trajectory effectively passes
directly from the reactant well to product well without entering a well corresponding
to an `intermediate' species.

It is natural to ask whether the above qualitative distinction between sequential and concerted reaction
pathways \cite{Carpenter04,Bachrach07}, made on the basis of considerations of the nature of
isomerizing trajectories in configuration space, can be given a rigorous
dynamical formulation in phase space.  In the present work we provide such a formulation, based
on the properties of the normal form in the vicinity of the index two saddle point.
In particular, we show that it is possible to define phase space dividing surfaces for higher index saddles
which generalize the now-familiar dividing surfaces defined for index one saddles
\cite{Wiggins90,wwju,ujpyw,WaalkensBurbanksWiggins04,WaalkensWiggins04,WaalkensBurbanksWigginsb04,
WaalkensBurbanksWiggins05,WaalkensBurbanksWiggins05c,SchubertWaalkensWiggins06,WaalkensSchubertWiggins08,
MacKay90,Komatsuzaki00,Komatsuzaki02,Wiesenfeld03,Wiesenfeld04,Wiesenfeld04a,Komatsuzaki05,Jaffe05,Wiesenfeld05,Gabern05,Gabern06,Shojiguchi08}
to the case of higher indices, and that phase points can be chosen on such
dividing surfaces with prescribed dynamical character (e.g., concerted crossing trajectories).

The structure of this paper is as follows.
In Sec.\ \ref{sec:ndof}  we review and extend our previous
analysis \cite{Ezra09} of the phase space structure in the vicinity of a
(non-resonant) index $k$ saddle in terms of the normal form.
Particular emphasis is given to discussion of the dynamical significance of
the values of the associated action integrals, and to
the symbolic representation of the qualitatively distinct classes of
trajectory behavior in the vicinity of the saddle-saddle equilibrium.
For generic non-resonance conditions on the
eigenvalues of the matrix associated with the linearization of Hamilton's equations about
the equilibrium point, the normal form Hamiltonian is integrable. Integrability
provides all of the advantages that separability provides for quadratic
Hamiltonians: the saddle dynamics can be described separately and the
integrals associated with the saddle DoFs can be used
to characterize completely the geometry of trajectories passing through a
neighborhood of the equilibrium point.
As for the case of index one saddles, normally hyperbolic invariant manifolds (NHIMs)
\cite{Wiggins90,Wiggins94}
associated with index two saddles
are an important phase space structure and we give a brief discussion of
their existence and the role they play in phase space transport in the vicinity of
index $k$ saddles, following our work in \cite{Ezra09}.

In Sec.\ \ref{sec:crossing} we introduce the concept of concerted crossing trajectories.
The concerted crossing trajectories are defined in terms of the symbolic code introduced
previously \cite{Ezra09}, and realize the inuitive notion of direct, concerted passage
between wells via the hilltop region.  Those trajectories that are not concerted crossing
are potentially, but not necessarily, sequentially isomerizing trajectories.

Sec.\ \ref{sec:DS} defines the dividing surface (DS) for index $k$ saddles.  This is a key aspect of the present
work; we show that it is possible to define a codimension one surface in phase space
through which all concerted crossing trajectories must pass, and
which is everywhere transverse to the flow.  We also introduce a parametrization of the dividing surface.
This parametrization together with the use of the normal form enables  us to sample the
DS and select phase points on trajectories having
specified dynamical character.

In Section \ref{sec:model_potential} we present a numerical study
of the phase space structure in the vicinity of an index two saddle
in the context of a problem of chemical dynamics, namely, isomerization in a multiwell potential.
For isomerization on this model potential energy surface, which has multiple (four)
symmetry equivalent minima, analysis of the phase space structure
in the vicinity of the index two saddle enables a rigorous
distinction to be made between concerted crossing (``hilltop crossing'') isomerizing trajectories
and those trajectories that are not concerted (potentially sequentially isomerizing)
\cite{Carpenter04,Bachrach07}.
Our normal form based procedure for sampling the DS 
enables us to determine phase points lying on concerted crossing trajectories.
Numerical propagation forwards and backwards in time shows that such trajectories do indeed
have the properties intuitively associated with the concerted isomerizing pathway.
Sec.\ \ref{sec:summary} concludes.

\newpage
\section{The Poincar\'e-Birkhoff normal form in a phase space neighborhood of an index $k$ saddle}
\label{sec:ndof}

We begin by considering the normal form in the neighborhood of an equilibrium point of
saddle- $\ldots$-saddle-center-$\ldots$-center stability type,
where there are $k$ saddle degrees-of-freedom (DoF) and $n-k$ center degrees-of-freedom
(eq.\ \eqref{eq:saddle_center}).
We assume the usual non-resonance condition on the eigenvalues of the matrix associated
with the linearization of the Hamiltonian vector field about the equilibrium point (see, e.g. \cite{WaalkensSchubertWiggins08}). In particular,
we assume that the purely imaginary eigenvalues satisfy
the non-resonance condition $k_{k+1} \omega_{k+1} + \ldots + k_n \omega_n \ne 0$
for any $(n-k)$-vector of integers $(k_{k+1}, \ldots, k_n)$ with {\em not all}
the $k_i=0$ (that is, $(k_{k+1}, \ldots , k_n) \in {\mathbb Z}^{n-k}-\{0\}$)
and the real
eigenvalues satisfy the (independent) non-resonance condition
$k_1 \lambda_1 + \ldots +  k_k \lambda_k\ne 0$
for any $k$-vector of integers $(k_1,  \ldots,  k_k)$ with {\em not all}
the $k_i=0$, $i=1, \ldots, k$.
In this case the normal form transformation
transforms the Hamiltonian to an
even order polynomial in the variables
\begin{subequations}
\label{ndof_ints}
\begin{align}
I_i & = q_i p_i, \, i=1, \ldots, k, \\
I_j & = \frac{1}{2} \left( q_j^2 + p_j^2 \right), \, j=k+1, \ldots, n.
\end{align}
\end{subequations}
In other words, we can express the normal form Hamiltonian as:
\begin{equation}
H(I_1, I_2, I_3, \ldots, I_n),
\label{nf_ham}
\end{equation}
with associated Hamilton's equations:
\begin{subequations}
\label{nf_hameq}
\begin{align}
\dot{q}_i& =  \frac{\partial H}{\partial p_i}=
\frac{\partial H}{\partial I_i} \frac{\partial I_i}{\partial p_i} =  \Lambda_i q_i,  \\
\dot{p}_i& =  -\frac{\partial H}{\partial q_i}
= -\frac{\partial H}{\partial I_i}\frac{\partial I_i}{\partial q_i}= -\Lambda_i p_i, \qquad i=1, \ldots, k,\\
\dot{q}_j & =  \frac{\partial H}{\partial p_j}= \frac{\partial H}{\partial I_j}\frac{\partial I_j}{\partial p_j} =
\Omega_j p_j,  \\
\dot{p}_j & =  -\frac{\partial H}{\partial q_j}= -\frac{\partial H}{\partial I_j}\frac{\partial I_k}{\partial q_j}=
- \Omega_j q_j,
\qquad j=k+1, \ldots, n,
\end{align}
\end{subequations}
where we have defined the frequencies
\begin{subequations}
\begin{align}
\Lambda_i(\mathcal{I}) & \equiv \frac{\partial H}{\partial I_i}(\mathcal{I}), \;\; i=1,\ldots, k,\\
\Omega_j(\mathcal{I}) & \equiv \frac{\partial H}{\partial I_j}(\mathcal{I}), \;\; j=k+1,\ldots, n,
\end{align}
\end{subequations}
where $\mathcal{I} = (I_1, I_2, I_3, \ldots, I_n)$
and it can be verified by a direct calculation that the $I_j$, $j=1, \ldots, n$, are
integrals of the motion for \eqref{nf_hameq}.

The integrability of the normal form equations \eqref{nf_hameq} provides us
with a very straightforward way of characterizing the dynamics in a neighborhood of the
index $k$ saddle. The coordinates $(q_i, p_i)$, $i=1, \ldots, k$, describe
saddle-type or `reaction dynamics' ,
and the dynamics in the $k$ saddle planes can be further characterized by
the  {\em saddle integrals}, $I_i$, $i=1, \ldots, k$.
The coordinates $(q_i, p_i)$, $i=k+1, \ldots, n$, describe bounded motions (center-type dynamics)
or `bath modes', which can be further characterized by integrals $I_i$, $i=k+1, \ldots, n$.

We now introduce a canonical transformation of the saddle variables \cite{Ezra09}.
Passage of  trajectories over the
saddle  is more naturally described in terms of these new coordinates,
and computation of codimension one dividing surfaces for index $k$ saddles is also facilitated.
The transformation is given by:
\begin{subequations}
\label{trans1}
\begin{align}
q_i & =  \frac{1}{\sqrt{2}} \left( \bar{q}_i + \bar{p}_i \right), \\
p_i & =  \frac{1}{\sqrt{2}} \left( \bar{p}_i - \bar{q}_i \right), \quad i=1, \ldots, k,
\end{align}
\end{subequations}
with inverse
\begin{subequations}
\label{inv_trans1}
\begin{align}
\bar{q}_i & =  \frac{1}{\sqrt{2}} \left(  q_i - p_i \right), \\
\bar{p}_i & =  \frac{1}{\sqrt{2}} \left( p_i + q_i \right), \quad i=1,  \ldots k.
\end{align}
\end{subequations}
The transformation of variables given by eq.\ \eqref{inv_trans1},
where $i=1, \ldots, k$, and
the remaining variables are transformed by the identity transformation,
is canonical.  The variables $\bar{q}_i$, $i=1,\ldots,k$, are naturally identified with
physical configuration space coordinates in the vicinity of the saddle.
The Hamiltonian is given by eq.\  \eqref{nf_ham}, with action variables
\begin{subequations}
\label{ndof_ints_2}
\begin{align}
I_i & = q_i p_i  = \frac{1}{2} \left( \bar{p}_i^2 - \bar{q}_i^2 \right), \, i =1, \ldots, k,\\
I_j & = \frac{1}{2} \left( q_j^2 + p_j^2 \right), \, j=k+1, \ldots, n,
\end{align}
\end{subequations}
and Hamilton's equations then take the following form:
\begin{subequations}
\label{nf_hameq_2}
\begin{align}
\dot{\bar{q}}_i & =  \frac{\partial H}{\partial \bar{p}_i}
= \frac{\partial H}{\partial I_i}\frac{\partial I_i}{\partial \bar{p}_i}
= \Lambda_i \bar{p}_i,  \\
\dot{\bar{p}}_i & =  -\frac{\partial H}{\partial \bar{q}_i}= -\frac{\partial H}{\partial I_i}\frac{\partial I_i}{\partial \bar{q}_i}
= - \Lambda_i \bar{q}_i, \qquad i=1, \ldots, k,\\
\dot{q}_j & =  \frac{\partial H}{\partial p_j}= \frac{\partial H}{\partial I_j}\frac{\partial I_j}{\partial p_j}
=   \Omega_j p_j,  \\
\dot{p}_j & =  -\frac{\partial H}{\partial q_j}= -\frac{\partial H}{\partial I_j}\frac{\partial I_j}{\partial q_j}
=  - \Omega_j q_j,
\qquad j=k+1, \ldots, n,
\end{align}
\end{subequations}

\subsection{A Normally Hyperbolic Invariant Manifold}
\label{sec:NHIM}

As noted  previously \cite{Ezra09}, for $n>k$ the $2n-2k-1$ dimensional surface:
\begin{equation}
\calM \equiv \left\{ \bar{q}_1=\bar{p}_1=  \cdots =\bar{q}_k = \bar{p}_k=0, \, H(0, \ldots, 0, I_{k+1}, \ldots, I_n ) =E>0 \right\},
\label{eq:NHIM}
\end{equation}
is a normally hyperbolic invariant manifold (NHIM) in the energy surface $H(I_1, \ldots, I_n)=E>0$.
Moreover, this  NHIM has $2n-k-1$ dimensional stable and unstable
manifolds (within the fixed energy surface). Note that these stable and unstable manifolds
are codimension one in the energy surface only for $k=1$, i.e., index one
saddles.  Nevertheless, in the construction of dividing surfaces for index $k$ saddles
we will see that the NHIM (but not its stable and unstable manifolds) plays a similar
role for all $k$.

The case $n=k$, i.e. a $n$ degree-of-freedom system with an index $n$ saddle requires special consideration. In this case the NHIM corresponds to an equilibrium point and exists {\em only} on the $E=0$ energy surface. Thus, to get a non-trivial NHIM existing for $E>0$ we must have $n>k$.

 \subsection{Accuracy of the Normal Form}
 \label{sec:accuracy}

Here we briefly address some of the issues associated with the accuracy of the normal form.
The transformation to normal form is implemented by an algorithm that
operates in an iterative fashion by simplifying
 terms in the Taylor expansion about the equilibrium point order by order,
 i.e., the order $M$ terms are normalized, then the order $M+1$ terms are
 normalized, etc \cite{WaalkensSchubertWiggins08}. The algorithm is such that normalization at order $M$ does
 not affect any of the lower order terms (which have already been normalized).
 The point here is that although the algorithm can be carried out to
 arbitrarily high order, in practice we must stop the normalization
 (i.e., truncate the Hamiltonian) at some order $M$, after which we
make a restriction to some neighborhood of the saddle in which the resulting computations achieve
some desired accuracy. It is therefore necessary to determine the
accuracy of the normal form as a power series expansion truncated
at order $M$ in a neighborhood of the equilibrium point by comparing
the dynamics associated with
the normal form to the dynamics of the original system. There are several independent tests
that can be carried out to verify accuracy of the normal form, such as the following:
\begin{itemize}

\item Examine the extent to which integrals associated with the normal form
are conserved along
trajectories of the full Hamiltonian
(the integrals will be constant on trajectories of the normal form Hamiltonian).

\item Check invariance of specific invariant manifolds
(e.g.,  the NHIM,  its stable and unstable manifolds, the energy surface)
under dynamics determined by the full Hamiltonian.

\end{itemize}

Both of these tests require us to use the transformation between
the original coordinates and the normal form coordinates.
Software for computing the normal form as
well as the transformation (and inverse transformation) between the original
coordinates and the normal form coordinates can be downloaded from
\url{http://lacms.maths.bris.ac.uk/publications/software/index.html}.
Specific examples where  the accuracy of the normal form and its relation to $M$,
the fixed neighborhood of
the saddle, and the constancy
of integrals of the truncated normal form on trajectories of the full Hamiltonian
can be found
in refs
\onlinecite{WaalkensBurbanksWiggins04,WaalkensBurbanksWigginsb04,WaalkensBurbanksWiggins05,WaalkensBurbanksWiggins05b,WaalkensBurbanksWiggins05c}.
A general
discussion of accuracy of the normal form can be found in
ref.\ \onlinecite{WaalkensSchubertWiggins08}.

\newpage
\section{Crossing and concerted crossing trajectories}
\label{sec:crossing}

In this section we  introduce the notion of {crossing} trajectories.
\emph{Crossing} trajectories pass through
a neighborhood of the index $k$ saddle in such a way that \emph{all} the saddle
coordinates $\bar{q}_i$ change sign, and constitute a
dynamically well-defined subset of trajectories.
Our treatment of crossing trajectories
provides an excellent illustration of the power of the normal form (NF) and
the utility of the integrals of the motion in analyzing dynamics
in the vicinity of the saddle.

The definition of crossing trajectories given here is purely local, 
in that it relies on the values of the integrals computed using the NF in the vicnity of the saddle.
In our discussion of the isomerization dynamics
for a model multi-well system given below in Sec.\ \ref{sec:model_potential}, 
we establish a connection between the local crossing property
and the more global mechanistic notion of \emph{concerted} (as opposed
to sequential) isomerization trajectories; in this context, 
the crossing trajectories defined here are usefully described as
\emph{concerted crossing} (CC) trajectories.

In the normal form coordinate system the  `crossing' of a saddle  is analyzed by considering
only the saddle degrees-of-freedom, $(\bar{q}_i, \bar{p}_i)$, $i=1, \ldots, k$,
since the remaining coordinates remain bounded.
Crossing occurs when all coordinates $\bar{q}_i$,  $i=1, \ldots, k$, change sign
as the trajectory passes through a neighborhood of the index
$k$ saddle.
Whether or not a given coordinate $\bar{q}_i$ changes sign depends on
the value of the integral $I_i$.
If $\bar{q}_i$ changes sign in the vicinity of the saddle then $I_i >0$. If $I_i =0$ then $\bar{q}_i$
is zero or evolves to zero either in forward or negative time (depending on the initial condition).
We consider the boundary of the region of crossing trajectories, characterised by $I_i =0$
in more detail below.

More precisely, crossing and other trajectories are characterized as follows:
\begin{enumerate}

\item If $I_i > 0$ for \emph{all} $i=1, \ldots , k$,
then the trajectory is a crossing trajectory.

\item If $I_i = 0$ and $\bar{q}_i = 0$ for \emph{all} $i=1, \ldots , k$,
then the trajectory is on the NHIM and is not a crossing trajectory.

\item If $I_i = 0$ for any $i=1, \ldots , k$ and $\bar{q}_i \neq 0$ for
any point on the trajectory then $\bar{q}_i \neq 0$ for all points on the
trajectory (for finite time); the coordinate $\bar{q}_i$ therefore does not
change sign, and the trajectory is not a crossing trajectory.

\end{enumerate}

The signs of the integrals provide a rather coarse descriptor for crossing trajectories.
In ref.\ \onlinecite{Ezra09} a symbolic description of saddle crossing trajectories for index $k=2$ saddles
was introduced based on the sign change of the $\bar{q}_i$.
This symbolic description distinguishes all qualitatively distinct classes of
crossing trajectories. Here we
recall the discussion of the symbolic description of crossing trajectories given in
ref.\ \onlinecite{Ezra09}.

Consider the index $k=2$ case.
The symbolic description of the behavior
of a trajectory as it passes through a neighborhood of the index 2 saddle with respect
to the  coordinates $\qb_k, \, k=1, 2$, is expressed by the following four symbols,
$(f_1 f_2; i_1 i_2)$,
where $i_1 = \pm$, $i_2 = \pm$, $f_1 = \pm$, $f_2 = \pm$.
Here $i_k$, $k=1,2$, refers to the ``initial'' sign of $\qb_k$
as it enters the neighborhood of the index 2 saddle and
$f_k$, $k=1,2$, refer to the ``final'' sign of $\qb_k$,
as it leaves the neighborhood of the index 2 saddle.
For example, trajectories of type $(--;+-)$ pass over the
barrier from $\qb_1 >0$ to $\qb_1 <0$, but remain on the side
of the barrier with $\qb_2 <0$.

Based on the number of distinct sequences of length four of $+$ and $-$, there are $2^4=16$
qualitatively distinct classes of trajectory, However, there are only four types
of trajectories for which there is a change of
sign of {\em both}  coordinates $\qb_1$ and $\qb_2$ as they pass through a neighborhood of
the index 2 saddle, and these have symbolic descriptions $(++;--)$,
$(-+;+-)$, $(+-;-+)$ and $(--;++)$. We will see that this symbolic description of crossing
trajectories can be  directly related to the geometry of the dividing surface.

As noted previously \cite{Ezra09}, codimension one surfaces separate the different types
of trajectory.  For index 2 saddles these are the codimension one invariant manifolds.
given by $\qb_1=\pb_1$, $\qb_2 =\pb_2$ (i.e., $I_1 =0$, $I_2 =0$).
For example, the codimension one surface $\qb_1=\pb_1$
forms the boundary between trajectories of type $(++; +-)$ and
$(-+; +-)$, and so on.

An obvious generalization of this symbolic construction can be carried out for
the case of index $k$ saddles, $k >2$, where there are
$2^{2k}$ possible classes of trajectory.

\newpage
\section{The dividing surface associated with index $k$ saddles}
\label{sec:DS}

An essential component of the analysis of reaction dynamics from the phase space
perspective is the construction of a dividing surface (DS) through which all
reactant trajectories must pass, and which (locally) has the essential no-recrossing
property \cite{WaalkensSchubertWiggins08}.   In this section we
discuss construction of a codimension one (within the energy surface) DS for
an index $k$ saddle using the equations of motion  \eqref{nf_hameq_2}
associated with the normal form Hamiltonian of Sec.\ \ref{sec:ndof}.
This construction again demonstrates the utility of an analysis in terms of 
the `physical' saddle coordinates $\bar{q}_i$.
In addition to constructing the DS, using the normal form we are able to
(locally) classify reactive trajectories, thereby obtaining a rigorous phase
space characterization of the subset of barrier crossing trajectories
associated with concerted isomerization dynamics.

\subsection{Definition of dividing surface in the general case}
\label{sec:index_k}

We define a measure of the distance from the origin
(i.e., the saddle) in the configuration space of the saddle degrees-of-freedom as:
\begin{equation}
\calD 
\equiv  \frac{1}{2} \sum_{i=1}^{k} \bar{q}_i^2.
\label{eq:distance}
\end{equation}
The dividing surface we define contains the set of phase space points 
corresponding to the minimum distance from the
origin attained by crossing trajectories.  At any turning point in the variable $\calD$ we have
\begin{equation}
\dot{\calD} = \sum_{i=1}^{k} \bar{q}_i  \dot{\bar{q}}_i =  \sum_{i=1}^{k} \Lambda_i  \bar{q}_i \bar{p}_i =0.
\label{eq:min_1}
\end{equation}
This equation defines a critical point for $\calD$, but it is in fact a minimum since
\begin{equation}
\ddot{\calD} = \sum_{i=1}^{k} \Lambda_i  \left( \dot{\bar{q}}_i  \bar{p}_1 + \bar{q}_i  \dot{\bar{p}}_i  \right)
=  \sum_{i=1}^{k}  \Lambda_i^2 \left(  \bar{q}_i^2 +  \bar{p}_i^2  \right)\geq0,
\label{eq:min_2}
\end{equation}
and where we have used $\frac{d}{dt} \Lambda_i =0$ for dynamics under the 
normal form Hamiltonian.
Note that $\ddot{\calD}=0$  precisely when $\bar{q}_i = \bar{p}_i=0, \,  i=1, \ldots , k$, i.e.,
it is zero on the NHIM.

The dividing surface at constant $E$ is defined by the intersection of the
following two $2n-1$ dimensional surfaces:
\begin{subequations}
\label{eq:DS_1}
\begin{align}
S_1(\bar{q}_1,\bar{p}_1,  \ldots ,\bar{q}_k, \bar{p}_k, q_{k+1}, p_{k+1}, \ldots, q_n, p_n) & \equiv H(I_1, \ldots, I_n) - E  =0, \\
S_2(\bar{q}_1,\bar{p}_1,  \ldots ,\bar{q}_k, \bar{p}_k, q_{k+1}, p_{k+1}, \ldots, q_n, p_n) &
\equiv \sum_{i=1}^{k} \Lambda_i  \bar{q}_i \bar{p}_i =0.
\end{align}
\end{subequations}
Points on the DS lie on crossing trajectories and must therefore satisfy the additional conditions:

\begin{equation}
I_i > 0, \quad i=1, \ldots , k . \quad
\label{eq:constraint}
\end{equation}
The surface defined by \eqref{eq:DS_1} without the constraint in
\eqref{eq:constraint} includes trajectories which are not crossing
trajectories and we will refer to it as the \emph{extended dividing surface}. 
The boundary between the dividing surface and the rest of the extended dividing
surface consists of phase points satisying the 
conditions $I_1 =0$, and/or $I_2 =0$ and is codimension two, the
intersection of \eqref{eq:DS_1} with the codimension one invariant manifolds
$I_1=0$, $I_2=0$.  

It should be clear that the surface defined by \eqref{eq:DS_1} is codimension one,
and is restricted to the energy surface $H(I_1, \ldots, I_n) = E$ by construction.
However, we need to prove that it has the properties of a DS, that is,

\begin{enumerate}

\item {\bf The vector field is transverse to the DS.}

The DS is codimension two in the full phase space (and codimension one restricted to the energy surface)
and therefore has two vectors transverse to it at each point.

The Hamiltonian vector field $\bv_H$ associated with the normal form Hamiltonian $H$ is:
\begin{equation}
\bv_H = \left(\Lambda_1 \bar{p}_1,  \Lambda_1 \bar{q}_1, \ldots \Lambda_k \bar{p}_k,  \Lambda_k \bar{q}_k,
\Omega_{k+1} p_{k+1} ,  -  \Omega_{k+1} q_{k+1} , \ldots, \Omega_n p_n,  -  \Omega_n q_n \right).
\label{eq:vf}
\end{equation}
We will show that the Hamiltonian vector field $\bv_H$ is transverse
to the DS on the energy surface $S_1=0$.

The rate of change of $S_1$ along $\bv_H$ is necessarily zero:
\begin{equation}
dS_1(\bv_H) = \{H, H\} =0,
\end{equation}
where $\{ \cdot, \cdot\}$ denotes the usual Poisson bracket.
The rate of change of $S_2$ along $\bv_H$ is
\begin{subequations}
\begin{align}
dS_2(\bv_H) & = \{S_2, H\}  = \frac{d}{d t} S_2 \\
& = \sum_{i=1}^{k}  \Lambda_i^2 \left(  \bar{q}_i^2 +  \bar{p}_i^2  \right).
\end{align}
\end{subequations}
from \eqref{eq:min_2}.
Clearly, this expression is greater than or equal to zero.
It is zero precisely when $\bar{q}_j = \bar{p}_j=0, \,  j=1, \ldots , k$, i.e. it is zero on the NHIM.

\item {\bf All crossing trajectories pass through the dividing surface.}

It is evident that all trajectories of \eqref{nf_hameq_2}
that enter and leave a neighborhood of the origin necessarily
achieve such a minimum distance from the origin.
In particular, this is true for all trajectories satisfying $I_i > 0, \,  i=1, \ldots, k$.
Therefore all
crossing trajectories pass through the dividing surface.

\end{enumerate}

\subsection{Index 1 Saddles}
\label{sec:index1}

It is instructive to see how our construction of a dividing surface for index $k$ saddles reduces to
the familiar case of index $1$ saddles \cite{wwju, ujpyw}.  In this
case conditions \eqref{eq:DS_1} become
\begin{subequations}
\label{eq:DS_1_ind1}
\begin{align}
H(I_1, \ldots, I_n) & = E \\
  \bar{q}_1\bar{p}_1&  =0,
  \end{align}
  \end{subequations}
where points on the dividing surface are also subject to the additional constraint:
\begin{equation}
I_1 = \frac{1}{2} \left( \bar{p}_1^2 - \bar{q}_1^2\right)>0,
\label{eq:constraint_ind1}
\end{equation}
Now, in order to have $\bar{q}_1\bar{p}_1 =0$ {\em and}
$I_1 = \frac{1}{2} \left( \bar{p}_1^2 - \bar{q}_1^2\right)>0$ we must have
\begin{equation}
\bar{q}_1=0,
\end{equation}
which is the DS for index 1 saddles expressed in terms of the normal form coordinates $\qb$.

\subsection{Explicit parametrization of the DS for quadratic Hamiltonians}
\label{sec:param}

For the case of a quadratic Hamiltonian it is possible to give an explicit
parametrization of the dividing surface.
For simplicity we consider the case of an index 2 saddle for a system with 2 DoF.
For the two degree-of-freedom index 2 saddle the dividing
surface is 2 dimensional in the 3 dimensional energy surface, and visualization
of the dividing surface is therefore possible.
Such visualization provides insight into the
crossing dynamics for an index 2 saddle, and we therefore discuss
this topic in some detail. Moreover, we will see that the dividing
surface in the quadratic Hamiltonian case plays an important role  in
our algorithm for sampling the dividing surface associated with the full
normal form for general (non-quadratic) Hamiltonians.

The discussion given below is in the spirit of ref.\ \onlinecite{ww10}, where the geometry
associated with the quadratic Hamiltonian was used to to examine several different approaches
to visualizing the phase space structures that govern reaction dynamics for index 1 saddles.
A quadratic approximation to the Hamiltonian is expected to accurately capture
the relevant geometrical features for energies `close' to the energy of the saddle.

We examine an $n$ DoF quadratic Hamiltonian with an index 2 saddle as motivation
for the parameterisation we will develop but the value of the index $k$ in our formulas 
is kept general for use later.
In the quadratic case, $\Lambda_i  \equiv \lambda_i =\mbox{constant}$,
$\Omega_i  \equiv \omega_i =\mbox{constant}$,   and the Hamiltonian \eqref{nf_ham} reduces to:
\begin{subequations}
\begin{align}
H(I_1, I_2, I_3, \ldots, I_n) &= \sum_{i=1}^{k}  \lambda_i I_i  + \sum_{j=k+1}^{n}  \omega_j I_j \\
& = \sum_{i=1}^{k} \frac{1}{2} \lambda_i  \left( \bar{p}_i^2 - \bar{q}_i^2 \right)
+ \sum_{j=k+1}^{n} \frac{1}{2} \omega_j \left( q_j^2 + p_j^2 \right)
\end{align}
\end{subequations}
which defines the energy surface:
\begin{equation}
H(I_1, I_2, I_3, \ldots, I_n) = E,
\label{energy_phys_1_fnf}
\end{equation}
where we will consider the case $E>0$.
We define the energy $E_s$ associated with the saddle DoF to be 
\begin{equation}
E_{s} = \sum_{i=1}^{k} \frac{1}{2}\lambda_i  \left( \bar{p}_i^2 - \bar{q}_i^2 \right)
= E - \sum_{j=k+1}^{n} \frac{1}{2} \omega_j  \left( q_j^2 + p_j^2 \right).
\label{energy_phys_2_fnf}
\end{equation}
($\lambda_i$ and $\omega_i$ constants) and rewrite this equation as follows:
\begin{subequations}
\begin{align}
E_{s} +  \sum_{i=1}^{k} \frac{1}{2}\lambda_i \bar{q}_i^2 & =
\sum_{i=1}^{k} \frac{1}{2}\lambda_i \bar{p}_i^2 \\
& = 
E + \sum_{i=1}^{k} \frac{1}{2}\lambda_i \bar{q}_i^2  
- \sum_{j=k+1}^{n} \frac{1}{2} \omega_j  \left( q_j^2 + p_j^2 \right)
\end{align}
\end{subequations}

Two ellipsoids are defined in the phase space $(\bar{q}_i, \bar{p}_i)$, $i=1, \ldots , k$ as follows.
For some  parameter $R \geq 0$ the condition
\begin{equation}
R =  \sum_{i=1}^{k} \frac{1}{2}\lambda_i \bar{q}_i^2,
\label{R_eq_fnf}
\end{equation}
defines a $(k-1)$-dimensional ellipsoidal  surface in the $\qb$ configuration space and
\begin{equation}
E_{s}+R =  \sum_{i=1}^{k} \frac{1}{2}\lambda_i \bar{p}_i^2,
\label{E+R_eq_fnf}
\end{equation}
defines a $(k-1)$-dimensional ellipsoidal surface in the $\pb$ momentum space.

For the quadratic Hamiltonian,
the condition for a trajectory to achieve a
minimum distance from the origin in the $\qb$ configuration space is
(cf. \eqref{eq:min_1})
\begin{equation}
\dot{\calD} = \sum_{i=1}^{k} \bar{q}_i  \dot{\bar{q}}_i =
\sum_{i=1}^{k} \lambda_i  \bar{q}_i \bar{p}_i =0
\label{eq:r_dot_fnf}
\end{equation}

The case of an index 2 saddle with no center degrees-of-freedom
corresponds to a two degree-of-freedom quadratic Hamiltonian system having
an equilibrium point at the origin, where the matrix associated with  the
Hamiltonian vector field  has two pairs of purely real eigenvalues,
$\pm \lambda _1$, $\pm \lambda_2$, with $\lambda_1, \lambda_2 >0$.
The  two dimensional dividing surface in the three dimensional energy
surface given by $H=E_s=E$ is then given parametrically by:
\begin{subequations}
\label{DS_param_2D_fnf}
\begin{align}
(\bar{q}_1, \bar{q}_2) &= \sqrt{2R} \left( \frac{\sin \theta}{\sqrt{\lambda_1}},
\frac{\cos \theta}{\sqrt{\lambda_2}} \right), \\
(\bar{p}_1, \bar{p}_2) &= \pm \sqrt{2(E+R)} \left( \frac{\cos \theta}{\sqrt{\lambda_1}},
\frac{-\sin \theta}{\sqrt{\lambda_2}} \right)
\end{align}
\end{subequations}
where $0 \leq R \leq \infty$, $0 \leq \theta \leq 2 \pi$. It is straightforward
to check that points on surface \eqref{DS_param_2D_fnf} satisfy conditions
\eqref{R_eq_fnf}, \eqref{E+R_eq_fnf}, and \eqref{eq:r_dot_fnf},
and that the
Hamiltonian vector field is transverse to this surface.
However, as we have not yet imposed condition
\eqref{eq:constraint}, the extended dividing surface \eqref{DS_param_2D_fnf}
contains phase points in addition to those on crossing trajectories.

Projections of the full surface \eqref{DS_param_2D_fnf}
into various 3 dimensional subspaces of phase space are shown
in Figure \ref{fig:index2plot3d1},
without the constraints  on  $R$  and $\theta$ implied by the conditions \eqref{eq:constraint}.
Consequently, not all phase points on this surface lie on crossing trajectories, but the true
DS is embedded in it.
The parameter values  chosen are
$\lambda_1 =1$, $\lambda_2 = \sqrt{3}$, $E=1.0$, and we show
both signs of the square root in  \eqref{DS_param_2D_fnf}, with $0 \leq R \leq 1$.

Figures \ref{fig:index2plot3d1}a and \ref{fig:index2plot3d1}b show
$\bar{p}_1$ and $\bar{p}_2$, respectively, as functions of $\bar{q}_1$ and $\bar{q}_2$.
Note that $\bar{p}_1$, $\bar{p}_2$ are apparently continuous as
$\bar{p}_1$, $\bar{p}_2$ pass through zero. This is however deceptive.
If we examine equation \eqref{DS_param_2D_fnf} we see that for any configuration
$(\bar{q}_1, \bar{q}_2)$, the two parts of the surface given by the
positive and negative square roots in the second equation must be distinct,
since from equation \eqref{E+R_eq_fnf} even when $R$ is zero the values of
$\bar{p}_1$, $\bar{p}_2$ cannot both be zero.
Hence, although in Figure
\ref{fig:index2plot3d1}a the values of $\bar{p}_1$ are continuous on the axis
$\bar{q}_2 = 0$, $\bar{p}_1 = 0$ the values of $\bar{p}_2$ are not continuous
as can be seen from Figure \ref{fig:index2plot3d1}b, so that the two surface
components are disjoint.

Also note that, following the discussion in Section \ref{sec:crossing},
there are codimension one surfaces separating the different types
of trajectory.  These are the four codimension one invariant
manifolds defined by particular values of the integrals and
given by $\bar{q}_1=\bar{p}_1$, $\bar{q}_1=-\bar{p}_1$,
$\bar{q}_2 =\bar{p}_2$, $\bar{q}_2 =-\bar{p}_2$ (i.e., $I_1 =0$, $I_2 =0$).
For example, the codimension one surface $\bar{q}_1=\bar{p}_1$
forms the boundary between trajectories of type $(++; +-)$ and
$(-+; +-)$, and so on.

Figure \ref{fig:index2plot3d2} shows only the parts of the
 dividing surface  satisfying the constraints given by equation \eqref{eq:constraint}.
 In other words, we only plot points on \eqref{DS_param_2D_fnf} for which  $I_1 > 0$, $I_2 > 0$.

 In order to obtain explicit constraints we substitute the
 expressions for the phase points in terms of $R$ and $\theta$  into the integrals to obtain
\begin{subequations}
\label{eq:i_1}
\begin{align}
I_1 & =  \frac{\bar{p}_1^2 - \bar{q}_1^2}{2}  , \\
& =  \frac{E}{\lambda_1} \cos^2 \theta + \frac{R}{\lambda_1}  \left(\cos^2 \theta - \sin^2 \theta \right),  \\
& =  \frac{(E+2R)}{\lambda_1} \cos^2 \theta - \frac{2R}{\lambda_1}
\end{align}
\end{subequations}
and
\begin{subequations}
\label{eq:i_2}
\begin{align}
I_2 & =  \frac{\bar{p}_2^2 - \bar{q}_2^2}{2}, \\
& = \frac{E}{\lambda_2} \sin^2 \theta + \frac{R}{\lambda_2}  \left(\sin^2 \theta - \cos^2 \theta \right),  \\
& =  \frac{(E+2R)}{\lambda_2} \sin^2 \theta - \frac{R}{\lambda_2}.
\end{align}
\end{subequations}
Imposing the conditions $I_1 > 0$ and $I_2 > 0$ we then obtain the constraint
\begin{equation}
\label{eq:constraint_1}
\frac{R}{E+2R} < \sin^2\theta < \frac{E+R}{E+2R}\, .
\end{equation}
Any point $(R, \theta)$ on the dividing surface satisfying condition \eqref{eq:constraint_1}
lies on a crossing trajectory.

Figures \ref{fig:index2plot3d2}a--\ref{fig:index2plot3d2}d show four almost disjoint,
apparently continuous components of the surface which meet only at
points $\bar{p}_1 = 0$, $\bar{p}_2 = 0$
where $R = 0$ ($\bar{q_1} = \bar{q_2} = 0$) and $\bar{p_1}$, $\bar{p_2}$ satisfy equation \eqref{E+R_eq_fnf}.

It is apparent from equation \eqref{DS_param_2D_fnf} that, for points on the DS with
$0 < \theta < \frac{\pi}{2}$ (quadrant $\bar{q}_1 > 0$, $\bar{q}_2 > 0$),
for the positive square root we have $\bar{p}_1 > 0$, $\bar{p}_2 < 0$ and
CC trajectories of the type $(+-;-+)$, while for the negative square root
we have $\bar{p}_1 < 0$, $\bar{p}_2 > 0$ and CC trajectories of
the type $(-+;+-)$.

Similarly for $ \pi < \theta < \frac{3 \pi}{2} $ with points in the
quadrant $\bar{q}_1 < 0$, $\bar{q}_2 < 0$, for the positive square root we
have $\bar{p}_1 < 0$, $\bar{p}_2 > 0$ and trajectories of type $(-+;+-)$,
while for the negative square root we have $\bar{p}_1 > 0$,
$\bar{p}_2 < 0$ and points corresponding to CC
trajectories of type $(+-;-+)$.
There are therefore four disjoint pieces of the dividing surface, two of
type $(+-;-+)$ and two of type $(-+;+-)$.

For the angle ranges $ \frac{\pi}{2} < \theta < \pi$ and $ \frac{3 \pi}{2} < \theta < 2 \pi$ there are
similarly four disjoint pieces of the dividing surface: two of the type
$(++;--)$ and two of type $(--;++)$.

Figures \ref{fig:index2plot3d11}a and \ref{fig:index2plot3d11}b show
$\bar{p}_1(\bar{q}_1, \bar{q}_2)$ and $\bar{p}_2(\bar{q}_1, \bar{q}_2)$, respectively,  for
the portion of the DS corresponding to CC trajectories of type $(++;--)$.

\subsection{Sampling the dividing surface}
\label{sec:samp}

Sampling the DS for an index 2 saddle in the NF coordinate
set and then using  the sample points as initial conditions
enables us to obtain directly concerted crossing trajectories of
particular dynamical (symbolic) type.  (See Sec.\ \ref{sec:model_potential}.)

Although we have obtained a  parametrization of the DS for the index 2 saddle for the
quadratic Hamiltonian case, we would like to
be able to sample the DS for general Hamiltonians.
Sampling the DS in the quadratic case is in principle straightforward: we need to sample
points parametrized by $(R, \theta)$ in eq.\
\eqref{DS_param_2D_fnf} and use the NF coordinate transformation to
convert these points to the original set of physical coordinates.  The resulting phase points can be
used as initial conditions for trajectory integration.

For the case of a general Hamiltonian, there are however several complications:
\begin{enumerate}

\item The frequencies $\Lambda_1$, $\Lambda_2$ in \eqref{DS_param_2D_fnf} are
no longer constant, but are in general functions of the phase space coordinates.

\item The  energy $E$ in \eqref{DS_param_2D_fnf} is in general a
  nonlinear function of the action variables.

\item The truncated NF coordinate transformations  and the associated Hamiltonian are approximate,
and become less accurate the farther we move from the saddle point.

\item In addition to sampling the saddle coordinates, for $n \geq 3$ DoF
it is necessary  to sample the center planes, essentially partitioning the total energy
  between the center and saddle degrees of freedom but subject to the constraint $H = E$.

\item To use the sampled points as initial conditions for trajectories in
microcanonical (constant $E$) simulations we require them to lie on the energy surface of the
 full Hamiltonian.

  \item Ultimately, one would like to sample the DS uniformly with respect to some
  prescribed probability density.

\end{enumerate}

For the full nonlinear normal form one of the difficulties is that we need to
sample the actions $I_i$, $ i=1, \ldots , k$, and $I_j$,   $j=k+1, \ldots, n$, subject to the
nonlinear energy constraint \eqref{energy_phys_1_fnf}.

First, consider the problem of sampling phase points in the center planes.
The phase space of the center DoF can be sampled uniformly, either in a rectangular
grid in physical coordinates $(\bar{q}_j, \bar{p}_j)$ or in action-angle
variables $(I_j\, \phi_j)$,
but we need to restrict the range of these variables
so that the constraints implied by the condition \eqref{energy_phys_1_fnf}
on the total energy are obeyed.

There are two possibilities.

First, with zero energy in the saddle planes ($I_i=0$, $i = 1, 2$), we could
calculate the allowed range for each phase space coordinate pair $(\bar{q}_j,
\bar{p}_j)$ or each action $I_j$ from \eqref{energy_phys_1_fnf}.
These values are however non-linearly interdependent and the
calculation generally involves an iterative numerical procedure such as Newton's method.

Alternatively, we can sample the center mode
phase space over a sufficiently large but fixed region and then simply accept
points for which $H(I_1, \ldots, I_n) < E$ with $I_i=0$ for $ i=1, \ldots, k$.

For each set of center DoF variables so obtained we need to sample points in the
saddle planes such that $H(I_1, \ldots, I_n) = E$.
For $k=1$ (index 1) there are no free saddle plane parameters.
For index $k=2$ we sample values for parameters $\theta$ and $R$.
It is then necessary to solve
\eqref{energy_phys_1_fnf} and \eqref{eq:r_dot_fnf} numerically for the values
of $(\bar{q}_i, \bar{p}_i)$.

For the general (non-quadratic) index 2 saddle we see that a point with given $(R, \theta)$ in
\eqref{DS_param_2D_fnf} lies (by construction) on the surface
$S_1$, but does not in general lie on $S_2$ (although it might approximately do so).
Such a phase point satisfies
\eqref{eq:min_1},  but \eqref{DS_param_2D_fnf} now does not include the
non\-linear (higher order) terms in actions $I_1 \ldots, I_n$ in an expansion
of $H$.
We define the energy correction $\Delta E$ via the relation 
\begin{subequations}
\begin{align}
H(I_1, I_2, I_3, \ldots, I_n) &= \sum_{i=1}^{k}  \Lambda_i I_i  +
\sum_{j=k+1}^{n}  \Omega_j I_j +  \Delta E \\
& = \sum_{i=1}^{k} \frac{1}{2} \Lambda_i  \left( \bar{p}_i^2 - \bar{q}_i^2 \right)
+ \sum_{j=k+1}^{n} \frac{1}{2} \Omega_j \left( q_j^2 + p_j^2 \right) +  \Delta E \,,
\end{align}
\end{subequations}
where $\Delta E $ includes the higher order terms from \eqref{DS_param_2D_fnf} but
will also include NF errors which are in general  functions of $\bar{q}_i, \bar{p}_i$.

We now postulate a generalised parameterisation of \eqref{DS_param_2D_fnf}
appropriate for an index 2 saddle with $n-2$ bath modes:
\begin{subequations}
\label{DS_param_kD_fnf}
\begin{align}
(\bar{q}_1, \bar{q}_2) &= \sqrt{2R} \left( \frac{\sin \theta}{\sqrt{\Lambda_1}},
\frac{\cos \theta}{\sqrt{\Lambda_2}} \right), \\
(\bar{p}_1, \bar{p}_2) &= \pm \sqrt{2(E_{s}+R)} \left( \frac{\cos \theta}{\sqrt{\Lambda_1}},
\frac{-\sin \theta}{\sqrt{\Lambda_2}} \right)
\end{align}
\end{subequations}
where $R \geq 0$, $0 \leq \theta \leq 2 \pi$.
Note that the $\Lambda_i$ are in general no longer constants so that
the relations \eqref{DS_param_kD_fnf} are implicit equations for $\bqb \equiv (\bar{q}_1, \ldots, \bar{q}_k)$ and $\bpb\equiv (\bar{p}_1, \ldots, \bar{p}_k)$.
It is a simple matter to check that phase points parametrized by
\eqref{DS_param_kD_fnf} satisfy generalised forms of \eqref{R_eq_fnf},
\eqref{E+R_eq_fnf}, and \eqref{eq:r_dot_fnf} (with the $\lambda$ replaced by $\Lambda$)  if we
note that $\sum_{i=1}^{k}  \Lambda_i I_i$ formally defines the saddle energy $E_{s}$,
with
\begin{equation}
E_{s} =  \sum_{i=1}^{k}  \Lambda_i I_i = H(I_1, I_2, I_3, \ldots, I_n) -
\sum_{j=k+1}^{n}  \Omega_j I_j -  \Delta E \,.
\end{equation}
Such phase points satisfy both equations \eqref{eq:min_1} and \eqref{eq:DS_1}
and so lie on $S_1$ and $S_2$.

For given parameters $R, \theta$ the phase space coordinates 
$(\bar{q}_i, \bar{p}_i)$ must be calculated in
combination with the $\Lambda_i, E_{s}$, which are non-linear functions of
the $(\bar{q}_i, \bar{p}_i)$.
While this is in general a difficult task, for
the calculations reported here, the required values can be found 
by an iterative procedure based on the quadratic approximation to the Hamiltonian.

Extension of eq.\ \eqref{DS_param_2D_fnf} to the
index $k$ saddle case with $n-k$ bath modes
is possible but cumbersome.  Although we do not go into details here, a similar
technique can be applied.

Once we have a parametrization of the DS, we can use it to
choose points on the surface and  integrate
associated trajectories forward and backward in time.
Quantitative calculation of, for example, fluxes requires sampling
the DS according to a prescribed density or properly weighting the samples.

\newpage
\section{Index-2 saddles: model potentials}
\label{sec:model_potential}

In this section we consider isomerization dynamics in a 2 DoF model potential
exhibiting an index 2 saddle.  Using the definition of the dividing surface
in the vicinity of the index 2 saddle discussed above, we are able to
sample phase points on  crossing trajectories via the normal
form and the associated symbolic code.  It is found that the 
trajectories defined in this
way have the dynamical attributes one would intuitively associate with 
trajectories following a concerted isomerization mechanism; in this section
we therefore refer to such trajectories as \emph{concerted crossing} (CC)
trajectories.

 \subsection{2 DoF 4 well model potential}

 We consider a non-separable  2 DoF 4-well potential of the form
  \begin{equation}
  \label{eq:4_well_pot_1}
  v(\Qb_1, \Qb_2) =  -\alpha \Qb_1^2 + \Qb_1^4 - \Qb_2^2 + \Qb_2^4 + \beta \Qb_1^2 \Qb_2^2.
  \end{equation}
In our numerical computations we use parameter values $\alpha =2$, $\beta = 0.4$.
Contours of the potential for these parameters are shown in Fig. \ref{plot_1a}.
Note that in this figure the
horizontal axis is $\Qb_2$ and the vertical axis is $\Qb_1$.

  The  associated Hamiltonian is:
  \begin{equation}
  H = \frac{\Pb_1^2}{2} + \frac{\Pb_2^2}{2} +  v(\Qb_1, \Qb_2),
  \label{ham_1a}
  \end{equation}
  with equations of motion:
  \begin{subequations}
  \label{hameq_1a}
  \begin{align}
  \dot{\Qb}_1 & =  \frac{\partial H}{\partial \Pb_1} =  \Pb_1,  \\
  \dot{\Qb}_2 & =  \frac{\partial H}{\partial \Pb_2} =  \Pb_2,  \\
  \dot{\Pb}_1 & =  -\frac{\partial H}{\partial \Qb_1} = 2 \alpha \Qb_1 - 4 \Qb_1^3 - 2 \beta \Qb_1 \Qb_2^2,  \\
  \dot{\Pb}_2 & =  -\frac{\partial H}{\partial \Qb_2} = 2 \Qb_2 - 4 \Qb_2^3 - 2 \beta \Qb_1^2 \Qb_2.
  \end{align}
  \end{subequations}

For the range of parameter  values of interest, the potential $v$ has the following
set of critical points:
\begin{enumerate}

\item  Minima (4)
\begin{equation}
(\Qb_1, \Qb_2) = \left(\pm \sqrt{\frac{2 \alpha - \beta}{4 - \beta^2}},
\pm \sqrt{\frac{2 - \alpha\beta}{4-\beta^2}} \right),
\;\;
v = -\frac{\alpha^2 +1 - \alpha\beta}{4-\beta^2}
\end{equation}

\item  Index-1 saddle (2)
\begin{equation}
(\Qb_1, \Qb_2) = \left(\pm \frac{\sqrt{\alpha}}{\sqrt{2}}, 0 \right),
\;\; v = -\frac{\alpha^2}{4}
\end{equation}

\item  Index-1 saddle (2)
\begin{equation}
(\Qb_1, \Qb_2) = \left(0, \pm -\frac{1}{\sqrt{2}} \right),
\;\; v = -\frac{1}{4}
\end{equation}

\item  Index-2 saddle (1)
\begin{equation}
(\Qb_1, \Qb_2) = \left(0, 0 \right),
\;\; v=0.
\end{equation}

\end{enumerate}

The basic problem of interest associated with a potential of the form
\eqref{eq:4_well_pot_1} concerns the nature of the isomerization
dynamics (i.e., well-to-well transformations).  In particular,
we wish to distinguish in a dynamically rigorous and useful
way between `concerted' and `sequential' isomerization processes.

Taking $\alpha > \beta/2$ for definiteness, we have the following rough classification
of dynamics as a function of energy:
\begin{itemize}

\item  Confined regime:
\begin{equation}
 -\frac{\alpha^2 +1 - \alpha\beta}{4-\beta^2} \leq E \leq -\frac{\alpha^2}{4}
\end{equation}
Trajectories are trapped in the vicinity of
one of 4 possible  minima, and no isomerization is possible.

\item    Restricted isomerization:
\begin{equation}
-\frac{\alpha^2}{4} < E \leq \frac{1}{4}
\end{equation}
Trajectories can pass between pairs of wells connected by the lowest energy index 1 saddles.
Passage between wells connected by higher energy  saddles is not possible.

In this case the standard phase space picture can be applied to analyze isomerizations,
with reactant regions identified in the usual way.
There are 2 symmetry related NHIMs, and for each isomerization
reaction either a 2-state (RRKM) or 3-state (Gray-Rice \cite{Gray87,Rice96})
model can be used to describe the isomerization kinetics for each pair of wells.

\item  Unrestricted isomerization:
\begin{equation}
-\frac{1}{4} <E < 0.
\end{equation}
Trajectories have sufficient energy to pass from any well to any well,
but do not have enough energy to reach the index-2 saddle.

In this regime,  the only possible way for the system to pass from the lower left well
to the upper right well, say, is via  \emph{sequential} isomerization routes that proceed 
through either of the intervening wells (top left or lower right).
Concerted passage via hilltop crossing is not energetically feasible.

In this case there are 4 relevant NHIMs.  A standard RRKM model could presumably
be be used to analyze the kinetics (flux over saddles), or a 5-state generalized
Gray-Rice model can be applied \cite{Gray87,Rice96}.
In the latter case, there are 4 reactant regions (wells) with boundaries consising of
broken separatrices, and a 5th region lying outside the reactant region.
We do not pursue such an analysis here.

\item   ``Roaming/concerted crossing'' regime: $E \geq  0$.

Trajectories can wander freely over a
single connected region of configuration space that encompasses all four wells.
In this regime, a direct \emph{concerted} isomerization route
exists that connects, for example, the lower left and upper right
wells.  In principle this route coexists with the sequential routes discussed above; the
dynamical problem addressed here then concerns the possibility of making
a rigorous distinction between trajectories associated with
the concerted and sequential isomerization pathways.  This classification will be made
in phase space using the normal form computed in the vicinity of the index two saddle.

\end{itemize}

 \subsection{Sampling trajectories on the DS}

We wish to sample trajectories having prescribed
dynamical character (CC and non-CC) for the 2
DoF system with Hamiltonian \eqref{ham_1a} using the DS specified by
\eqref{eq:DS_1} with crossing trajectories subject to the constraint
\eqref{eq:constraint}.
The CC trajectories constitute a dynamically well-defined
subset of trajectories that are associated with concerted well-to-well transitions.
Isomerizing trajectories that enter the vicinity of an index $k$ saddle and which are
not crossing (non-CC) trajectories are potentially associated with
sequential well-to-well transformations.

The DS is sampled as
outlined in Section \ref{sec:samp}.  
A regular grid in cartesian
coordinates $(\xi = R \cos[\theta], \eta = R \sin[\theta])$
provides an associated set of $(R, \theta)$ values;
the quadratic frequencies $\lambda_1, \lambda_2$ are substituted
for the exact, nonlinear frequencies $\Lambda_1, \Lambda_2$ in
\eqref{DS_param_2D_fnf} as a first approximation to give a sample point in
the space of NF coordinates.  New values for $\Lambda_1, \Lambda_2$ are then computed from
\eqref{nf_hameq}, and simple iteration gives a point satisfying
\eqref{DS_param_2D_fnf}.

In general, the phase point thus obtained is not
on the energy surface, so we must solve numerically
for the value of $E_{s}$ for which the condition on the total energy 
is satisfied.
This is done by appropriate scaling of the momenta $\bar{p}$.
Each calculation of the energy error involves a separate iteration
to obtain the nonlinear frequencies $\Lambda_i$; for the calculations reported here
a single iteration proves sufficient to achieve convergence.

A further difficulty arises as a consequence of the inherent inaccuracy 
of the truncated NF
transformations. If we solve for phase points with fixed  NF energy  as defined by
\eqref{energy_phys_1_fnf} and then use the NF coordinate transformation to
map the point into the original physical coordinates, the 
resulting point obtained no longer
lies exactly on the energy surface defined by the original physical 
Hamiltonian; the associated error increases with distance from the saddle.
We must therefore determine the value of $E_{s}$ yielding a NF
point which, after transformation, lies on the energy surface defined by the
original Hamiltonian (expressed in the original physical coordinates). This
procedure yields  points which we use as initial conditions for integrating
trajectories of the full Hamiltonian in the original physical coordinates.

In Figures \ref{fig:cc_traj_0.01},\ \ref{fig:cc_traj_0.1},\
\ref{fig:cc_traj_0.5} we show trajectories in $(\bar{Q}_1, \bar{Q}_2)$ space 
obtained by sampling points on the DS as described above.
The spacing of the sampling grid in $(\xi, \eta)$ is set at 
$0.01$ with $R \leq 0.1$.
Our sampling procedure leads to a sparsity of 
points on the DS near the origin in configuration space, $(\Qb_1, \Qb_2)=(0,0)$. Note also
that $\theta$ is undefined at $R=0$.

The total energies for the 2 DoF system are set at $E = 0.01, 0.1, 0.5$, respectively.
As we do not impose condition \eqref{eq:constraint}, 
we sample the extended dividing surface and so
obtain points in addition to those on concerted crossing trajectories.

At each sample point a trajectory is
integrated forwards and backwards in time, classified by symbolic code (as
above) and plotted.
Each of the Figures \ref{fig:cc_traj_0.01}, \ref{fig:cc_traj_0.1} and 
\ref{fig:cc_traj_0.5} has 4 panels, where each panel shows trajectories associated with
points on the DS having the following symbolic classifications:
concerted crossing (classes $(++;--)$ and  $(+-;-+)$), and
non-CC trajectories (classes $(+-;--)$ and $(++;-+)$).

The results shown in Figs \ref{fig:cc_traj_0.01}--\ref{fig:cc_traj_0.5} are 
noteworthy in several respects. First, it is clear that, for
all three values of the energy considered, the symbolic classification of 
trajectories as obtained using the NF in the vicinity of the index-2 saddle
corresponds precisely to the observed 
dynamical behavior of the numerically integrated trajectories.
That is, the NF enables us to sample a subset of trajectories of a given dynamical type, e.g.,
concerted crossing.
Second, we note that, although the ensemble of CC trajectories for a given energy appears
to consist of two disjoint pencils or bundles, the set of initial consitions is in fact connected
(there is a CC trajectory passing through the origin). 
The form of the configuration space projections of the various trajectory
classes is by no means inuitively obvious: at all three energies studied, CC trajectories tend to
be concentrated away from the hilltop itself.  The non-CC trajectories passing through the 
vicinity of the index 2 saddle tend to `bounce' off the saddles as they pass between
the wells, while CC trajectories appear to `graze' the hilltop. 

As mentioned previously, the boundary between CC and non-CC trajectories on the extended
dividing surface is composed of those points on the DS satisfying $I_1=0$ and/or $I_2=0$.
Projected into configuration space, the boundary consists of a number of 
lines emanating from the origin.  
In Fig.\ \ref{fig:boundary_0.01} we show the projection into configuration
space of a segment of a boundary between CC and non-CC trajectories 
defined by the condition $I_2=0$ with $0 \leq R \leq 0.1$ and $E =0.01$.
Each phase point on the boundary is propagated forward and backward in time.
It can be seen that trajectories associated with boundary points start in the lower 
left hand well, pass through the vicinity of the saddle and end up (at short times)
tending towards the vertical axis, i.e., neither left (non-CC) nor right (CC).
Again, the phase points on the boundary are obtained using the NF, yet show 
precisely the expected dynamical behavior when propagated numerically.

Lastly, it is natural to consider the fraction of CC trajectories in a given ensemble at
a prescribed energy.  To provide an unambigous determination of this quantity, we 
must carefully define the relevant ensemble, such as we have done above in terms of
the sampling procedure on the DS, and also specify the relevant
weighting factor (measure) for trajectories.  As the coordinates $(R, \theta)$ used 
to parametrize the DS are not canonical coordinates, a weighting factor enters that is 
a non-trivial function of $(R, \theta)$.  Rather than calculate the fraction
of CC trajectories by sampling the DS using this weight function, in the next subsection
we consider a different sampling procedure where trajectories are initiated on
the plane $\bar{Q}_1 = 0$ with $\bar{P}_1 >0$.  In this case the
associated density at constant energy is just the natural measure (area)
in the $(\bar{Q}_2, \bar{P}_2)$ plane.

\subsection{Trajectory studies of isomerization dynamics}

We now study the isomerization dynamics in the concerted crossing regime 
using an approach complementary to that of the previous subsection. 
  These computations provide additional insight into the way in which the presence of the index-2
  saddle affects trajectories in the neighborhood of the saddle, and further confirm the
  accuracy of the NF in the vicinity of the index 2 saddle.

  Initial conditions are chosen as follows. In the configuration space $(\Qb_1, \Qb_2)$,
  we define a regular grid of points along the line $\Qb_1=0$ in an interval symmetric
  about $\bar{Q}_2 =0$.   At each point $(\Qb_1, \Qb_2)$ along this line,
  a number of momentum pairs $(\Pb_1, \Pb_2)$ are chosen 
  such that the phase space points  $(\Qb_1, \Qb_2, \Pb_1, \Pb_2)$ lie 
  on the fixed energy surface, $H=E$.
  We consider the same 3 energies as in the previous subsection, 
  $H(\Qb_1, \Qb_2, \Pb_1, \Pb_2)=0.01$, $0.1$ and $0.5$, and we set $\rmd {\Qb}_1/\rmd t \geq 0$, so that
  trajectories start out moving from the lower to the upper
  half of the potential (the potential is symmetric by construction).

  A symbolic code can be assigned to each initial condition in either of two ways.
  Trajectories can be integrated forward and backward in time until the first well
  is reached.   In practice this means that we integrate trajectories until a turning point
  is reached in either of the coordinates $\bar{Q}_{1,2}$.
  In forward time the trajectory can enter one of 2 possible wells, and in backward time the
  trajectory can enter 2 possible wells, so that there are 4
  qualitatively different type of trajectories.  Using our symbolic classification \cite{Ezra09}, 
  the trajectory  $(f_1, f_2; i_1, i_2)$,
  where  $i_k=\pm$, $f_k = \pm$, $k=1, 2$, denotes the trajectory
 that passes from well $(i_1, i_2)$ in the immediate past to
 the well $(f_1, f_2)$ in the immediate future.
 Each initial condition on the horizontal line can therefore 
 be labelled  according to its symbolic description $(f_1, f_2; i_1, i_2)$, i.e.,
 the first well visited in the past and future, as determined by the exact trajectory dynamics. 
 We can also use the NF to classify phase points in the $\bar{Q}_1=0$ plane, and 
 this symbolic classification can be compared with the trajectory results.

 Initial conditions at constant energy with $\Qb_1 =0$, $\Pb_1 \geq 0$
 are uniquely specified by the values of the phase space variables  $(\Qb_2, \Pb_2)$.
 We therefore assign the symbolic trajectory code (4 possibilities) as a function of
 coordinates $(\Qb_2, \Pb_2)$.  Initial conditions are sampled uniformly
 in the $(\bar{Q}_2, \bar{P}_2)$ plane.

 Our results are presented in Fig.\ \ref{fig:q2p2_f}
in which we plot the fraction $F$
 of crossing trajectories as a function of coordinate $\Qb_2$.
 Results are shown for three energies
 $E=0.01$, $0.1$ and $0.5$ across the interval $-0.5 \leq \Qb_2 \leq 0.5$. 
 This range of the coordinate $\bar{Q}_2$ corresponds approximately to the
 constraint $0 \le R \leq 0.1$ employed when sampling the DS as
 described in the previous subsection.
 On the same graph we show the corresponding 
 trajectory fractions 
 determined by taking each initial condition, converting it to NF coordinates,
 and using the NF to predict the symbolic trajectory code. 
In Figure \ref{fig:q2p2_f} the sample spacing for trajectory initialisation was
$\Delta \bar{Q}_2 = 0.01$,  while for the faster normal form sampling we used a 5 times finer 
spacing, $\Delta \bar{Q}_2 = 0.002$.  Use of the NF results in a smoother curve.
If the same sampling spacing is used as for the trajectory calculations, the NF 
results are essentially identical to those obtained by
trajectory integration.
 Discrepencies between the trajectory and NF results 
 are most pronounced near the boundaries of the interval, and this
 is not unexpected as the accuracy of the NF presumably deteriorates
 as one moves away from the saddle.  What is perhaps striking about our results
 is the size of the region of the $\bar{Q}_2$ axis over which the NF predictions 
 do accurately match the trajectory results.

 \subsection{Saddle crossing in the presence of bath modes}
 
 Finally, we briefly explore the nature of 
 the saddle crossing dynamics in the presence of additional degrees of freedom
 (bath modes). 
 We consider a 4 DoF model in which 2 bath modes are added to the 4-well 2 DoF 
 system studied above, and bilinear coupling terms are introduced between
 saddle and bath modes.

 \subsubsection{System-bath Hamiltonian}

 We consider a 4 DoF system with an index 2 saddle point.  The relevant
 system-bath  Hamiltonian $H_{\text{sb}}$ is obtained by adding two `bath' modes, coordinates
 $(x, y)$, to the
 2 DoF system with Hamiltonian given by eq.\ \eqref{ham_1a}:
   \begin{equation}
    \label{sb_ham}
  H_{\mbox{sb}}  =
  {\frac{{\Pb_1}^2}{2} +  \frac{{\Pb_2}^2}{2} +
  V(\Qb_1, \Qb_2)}
   +
  {\frac{1}{2} \left[ p_{x}^2 +
  \left(\omega_x x  -\frac{c_{1} \Qb_1}{\omega_x} \right)^2
  \right]}
  +
  { \frac{1}{2} \left[ p_{y}^2 + \left(
  \omega_y y - \frac{c_{2} \Qb_2}{\omega_x} \right)^2
   \right]},
  \end{equation}  
  where mode $x$ is coupled to system coordinate $\bar{Q}_1$ and mode $y$
  is coupled to system coordinate $\bar{Q}_2$.
  
  Hamilton's equations are given by:
    \begin{subequations}
    \label{hameq_3a}
    \begin{align}
    \dot{\Qb}_1 & =  \frac{\partial H}{\partial \Pb_1} =  \Pb_1,  \\
    \dot{\Qb}_2 & =  \frac{\partial H}{\partial \Pb_2} =  \Pb_2,  \\
    \dot{x} & = \frac{\partial H}{\partial p_x} = p_x,\\
    \dot{y} &=  \frac{\partial H}{\partial p_y} = p_y, \\
    \dot{\Pb}_1 & =  -\frac{\partial H}{\partial \Qb_1} = 2 \alpha \Qb_1 - 4 \Qb_1^3 - 2 \beta \Qb_1 \Qb_2^2 
    + \frac{c_1}{\omega_x} \left(\omega_x x
    -\frac{c_{1} \Qb_1}{\omega_x} \right),  \\
    \dot{\Pb}_2 & =  -\frac{\partial H}{\partial \Qb_2} = 2 \Qb_2 - 4 \Qb_2^3 - 2 \beta \Qb_1^2 \Qb_2 
    + \frac{c_2}{\omega_y} \left(
  \omega_y y - \frac{c_{2} \Qb_2}{\omega_y} \right), \\
    \dot{p}_x & = -\frac{\partial H}{\partial x} =-\omega_1\left(\omega_x x  -\frac{c_{1} \Qb_1}{\omega_x} \right),\\
    \dot{p}_y & = -\frac{\partial H}{\partial y}=-\omega_y \left(
  \omega_y y - \frac{c_{2} \Qb_2}{\omega_y} \right).
    \end{align}
    \end{subequations}
   For our numerical calculations we use bath parameters $ \omega_x = 1.0$, $\omega_y = \sqrt{2}$ and set
   $c_1 = c_2$.

 \subsubsection{Sampling trajectories on the extended DS for 4 DoF}

 We now consider the effect of additional degrees of freedom on the dynamics, in particular,
 our ability to sample concerted crossing trajectories on the (extended) DS for the 4 DoF
 system-bath Hamiltonian,  eq.\ \eqref{sb_ham}.
 
 The energy shell for the 4 DoF system-bath model is 7 dimensional, while the DS is 
 6 dimensional.  Fully sampling the DS according to the procedures described
 above is a numerically intensive task. 
 For purposes of illustration, we demonstrate that computation of the NF for
 the 4 DoF model allows us to sample CC trajectories on the DS, and that 
 bundles of such trajectories, when projected into the $(\bar{Q}_1, \bar{Q}_2)$ subspace,
 appear simply as `fattened' versions of the  2DoF saddle trajectories
 described above with an appropriate value of the saddle energy.  
 This inherent apparent simplicity of the system-bath dynamics is a consequence of the
 integrability of the NF in the vicinity of the saddle.

 In order to sample the DS,  we first set the saddle energy $E_s = 0.1$
 (no excitation in center modes) and sample phase space coordinates 
 in the saddle planes; for each saddle plane point we
 sample phase points in the center DoF setting $x = \pm 0.5$, $y = \pm 0.5$, $p_x = \pm 0.5$, and 
 $p_y = \pm 0.5$,  and scale the center DoF coordinates and momenta 
to obtain a fixed value of the total energy $E_{\text{sb}} = 0.5$.

Figure \ref{fig:ccb_traj_0.1_0.1} shows $(\bar{Q}_1, \bar{Q}_2)$ projections
of trajectories obtained using the
sampling procedure just described with saddle-bath coupling parameter $c_1 = c_2 = 0.1$.

Each initial condition in the saddle planes
is therefore associated with a `bundle' of trajectories for the 4 DoF full system.

Figure \ref{fig:ccb_traj_0.1_0.5}  shows analogous results for larger system-bath coupling 
parameters $ c_1 = c_2 = 0.5$.  For the larger coupling we obtain 
a thicker bundle associated with the fiducial initial condition in 
the saddle planes.

\newpage
\section{Summary and conclusions}
\label{sec:summary}

In this paper we have extended our earlier analysis of the phase space structure in the vicinity of
an equilibrium point associated with an index $k$ saddles \cite{Ezra09}.
We have shown that Poincar\'e-Birkhoff normal form theory provides a constructive procedure
for obtaining an integrable approximation to the full Hamiltonian in the vicinity of
the equilibrium, provided a generic non-resonance condition is satisfied independently 
for both the real eigenvalues and the complex eigenvalues of the matrix associated 
with the linearization of Hamilton's equations about the index $k$ saddle.  
As a consequence there are $k$ independent integrals associated with the 
saddle degrees-of-freedom. These integrals provide a precise tool for classifying 
trajectories that pass through a neighborhood of the saddle.  
In particular, they provide a symbolic classification of the trajectories 
into $2^{2k}$ distinct types of trajectory that pass through a neighborhood of the index $k$ saddle.

The normal form also provides an algorithm for constructing a dividing surface, i.e., 
a co-dimension one surface (restricted to the energy surface) through which all trajectories 
in a neighborhood of the index $k$ saddle must pass (with the exception of a set of zero measure). 
We provide a parametrization of this dividing surface which, when using the integrals associated 
with the  saddle degrees-of-freedom, can be sampled in such a way that we can 
choose initial conditions corresponding to any particular  
type of trajectory described by the symbolic classification.

We illustrated our analytical and computational techniques by analyzing a 
problem that brings to light a fundamental mechanistic role played by 
index two saddles in chemical dynamics. Namely, we consider isomerization 
on a potential energy surface with multiple
symmetry equivalent minima. In the two degree-of-freedom example we computed the normal form 
and the dividing surface and showed that the 
different classes of reactive trajectories in the
vicinity of the index two saddle (for three different energies)  
could be computed by our sampling routine. 
Our procedure enables a rigorous definition of concerted crossing
trajectories to be given in terms of local phase space structure. 
We then considered a  simplified 
system-bath model (one harmonic oscillator mode for each degree-of-freedom),
and showed that our approach could be applied to this four degree-of-freedom system.

\acknowledgments

PC and SW  acknowledge the support of the  Office of Naval Research Grant No.~N00014-01-1-0769.
PC, GSE, and SW  would like to
acknowledge the stimulating environment of the NSF sponsored Institute for
Mathematics and its Applications (IMA) at the University of Minnesota,
where this work  was begun.


\begin{thebibliography}{10}
\bibitem{saddle_footnote2}
\bibinfo{note}{{Without loss of generality, throughout this paper all
  equilibrium points considered will be located at the origin with zero
  energy.}}
\end{thebibliography}

\def\cprime{$'$}

\newpage

\section*{Figure captions}

\begin{figure}[H]
\caption{Schematic representation of sequential versus concerted isomerizing pathways in a model 4-well
potential. (a)  Sequential. (b) Concerted.
}
\label{fig:seq_con}
\end{figure}

\begin{figure}[H]
\caption{Projections of the
index 2 saddle phase space extended dividing surface, parameter values 
$\lambda_1 =1$, $\lambda_2 = \sqrt{3}$, $E=1.0$, with $0 \leq R \leq 1$.
(a) Coordinates ($\bar{q}_1$, $\bar{q}_2$, $\bar{p}_1$).
(b) Coordinates ($\bar{q}_1$, $\bar{q}_2$, $\bar{p}_2$).
(c) Coordinates ($\bar{p}_1$, $\bar{p}_2$, $\bar{q}_1$).
(d) Coordinates ($\bar{p}_1$, $\bar{p}_2$, $\bar{q}_2$).
}
\label{fig:index2plot3d1}
\end{figure}

\begin{figure}[H]
\caption{Projections of the
index 2 saddle phase space dividing surface with the constraints
\eqref{eq:constraint},
 parameter values $\lambda_1 =1$, $\lambda_2 = \sqrt{3}$, $E=1.0$.
(a) Coordinates ($\bar{q}_1$, $\bar{q}_2$, $\bar{p}_1$).
(b) Coordinates ($\bar{q}_1$, $\bar{q}_2$, $\bar{p}_2$).
(c) Coordinates ($\bar{p}_1$, $\bar{p}_2$, $\bar{q}_1$).
(d) Coordinates ($\bar{p}_1$, $\bar{p}_2$, $\bar{q}_2$).
}
\label{fig:index2plot3d2}
\end{figure}

\begin{figure}[H]
\caption{Projections of the portion of the index 2 saddle phase space dividing surface
associated with trajectories of symbolic type $++--$.
Parameter values $\lambda_1 =1$, $\lambda_2 = \sqrt{3}$, $E=1.0$.
(a) Coordinates ($\bar{q}_1$, $\bar{q}_2$, $\bar{p}_1$).
(b) Coordinates ($\bar{q}_1$, $\bar{q}_2$, $\bar{p}_2$).
}
\label{fig:index2plot3d11}
\end{figure}

  \begin{figure}[H]
 
  \caption{\label{plot_1a} Contour plot of nonseparable 2 DoF potential
  $v(\Qb_1, \Qb_2)$, eq.\ \eqref{eq:4_well_pot_1}.  Parameter values
  $\alpha = 2$, $\beta = 0.4$.}
  \end{figure}

   \begin{figure}[H]
    \caption{Trajectories initiated on the DS and propagated forwards and backwards in time.
   Saddle energy is $0.01$.
   (a) Concerted crossing trajectories $(++;--)$.
   (b) Concerted crossing trajectories $(+-;-+)$.
   (c) Non-CC trajectories $(+-;--)$, $I_2 <0$. 
   (d) Non-CC trajectories $(++;-+)$, $I_2 <0$. 
   \label{fig:cc_traj_0.01}
   }
  \end{figure}

    \begin{figure}[H]
       \caption{Trajectories initiated on the DS and propagated forwards and backwards in time.
       Saddle energy is $0.1$.
   (a) Concerted crossing trajectories $(++;--)$.
   (b) Concerted crossing trajectories $(+-;-+)$.
   (c) Non-CC trajectories $(+-;--)$, $I_2 <0$. 
   (d) Non-CC trajectories $(++;-+)$, $I_2 <0$. 
       \label{fig:cc_traj_0.1}
   }
  \end{figure}

   \begin{figure}[H]
     \caption{Trajectories initiated on the DS and propagated forwards and backwards in time.
   Saddle energy is $0.5$.
   (a) Concerted crossing trajectories $(++;--)$.
   (b) Concerted crossing trajectories $(+-;-+)$.
   (c) Non-CC trajectories $(+-;--)$, $I_2 <0$. 
   (d) Non-CC trajectories $(++;-+)$, $I_2 <0$. 
   \label{fig:cc_traj_0.5}
   }
 \end{figure}

 \begin{figure}[H]
  \caption{Segment of the boundary between CC and non-CC trajectories 
 defined by the condition $I_2=0$.  Crosses indicate configuration space
 projections of boundary points with $0 \leq R \leq 0.1$ and $E =0.01$.
 Each phase point on the boundary is propagated forward and backward in time.
 \label{fig:boundary_0.01}
 }
 \end{figure}

\begin{figure}[H]
\caption{Fraction $F$ of crossing trajectories  
for initial conditions at constant energy in the $\Qb_1 =0$ plane
with $-0.5 \leq \Qb_2 \leq +0.5$.  
NF predictions (green line) are show together with 
results obtained from integration of trajectories (red line).
(a) $E = 0.01$,  (b) $E = 0.1$, (c) $E = 0.5$.
\label{fig:q2p2_f}
}
\end{figure}

 \begin{figure}[H]
 \caption{Trajectories initiated on the DS and propagated forwards and backwards in time.
   Total energy $E = 0.1$.  Bath coupling parameter $c_1 = c_2 = 0.1$.
   (a) Concerted crossing trajectories $(++;--)$.
   (b) Concerted crossing trajectories $(+-;-+)$.
   (c) Non-CC trajectories $(+-;--)$, $I_2 <0$. 
   (d) Non-CC trajectories $(++;-+)$, $I_2 <0$. 
 \label{fig:ccb_traj_0.1_0.1}
 }
 \end{figure}

  \begin{figure}[H]
    \caption{Trajectories initiated on the DS and propagated forwards and backwards in time.
     Total energy $E = 0.1$.  Bath coupling parameter $c_1 = c_2 = 0.5$.
   (a) Concerted crossing trajectories $(++;--)$.
   (b) Concerted crossing trajectories $(+-;-+)$.
   (c) Non-CC trajectories $(+-;--)$, $I_2 <0$. 
   (d) Non-CC trajectories $(++;-+)$, $I_2 <0$. 
   \label{fig:ccb_traj_0.1_0.5}
 }
  \end{figure}


\newpage

\hskip -3.0cm
\includegraphics[width=20.0cm]{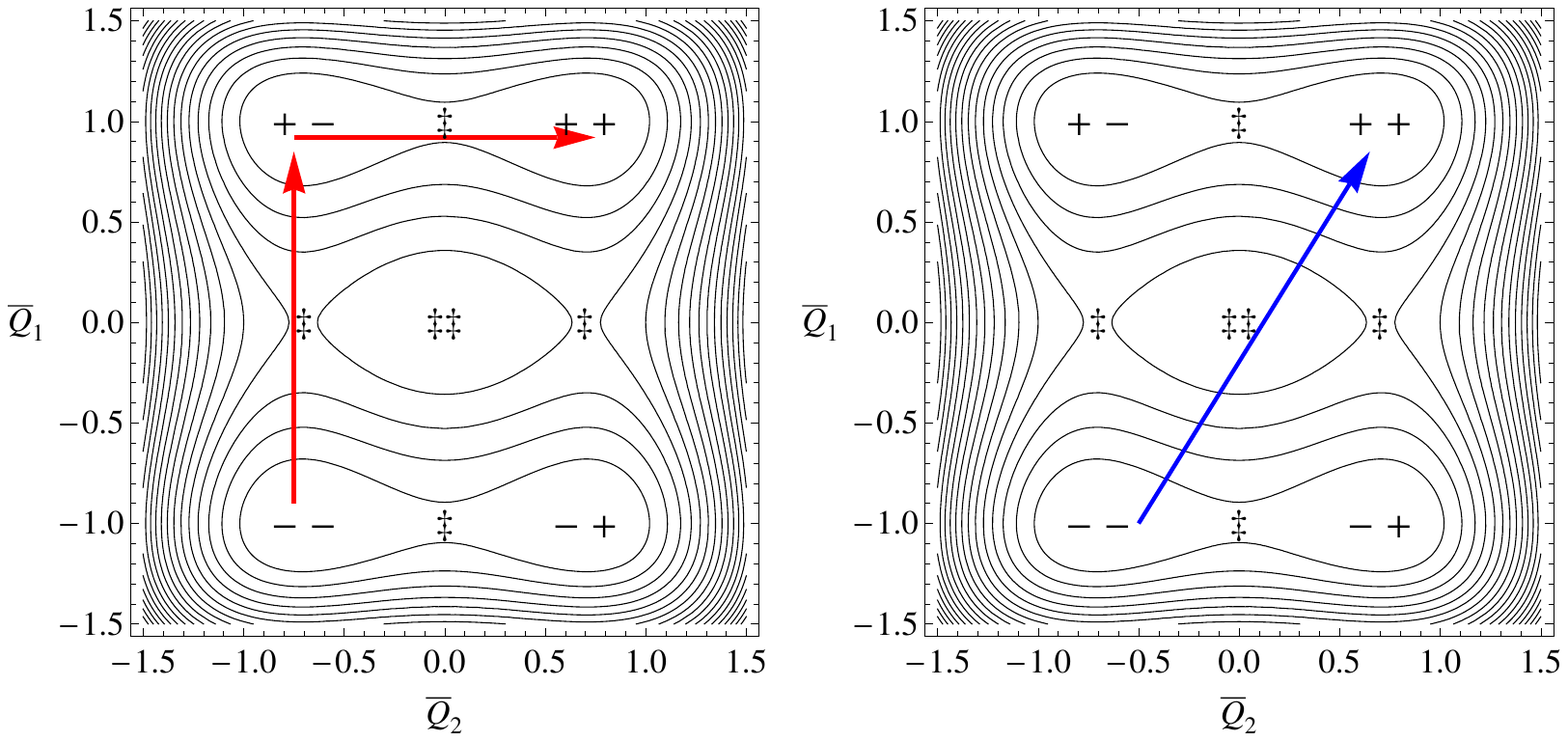}

 \vspace*{-6.5cm}
  FIGURE 1

\newpage

\hskip -3.0cm
\includegraphics[width=20.0cm]{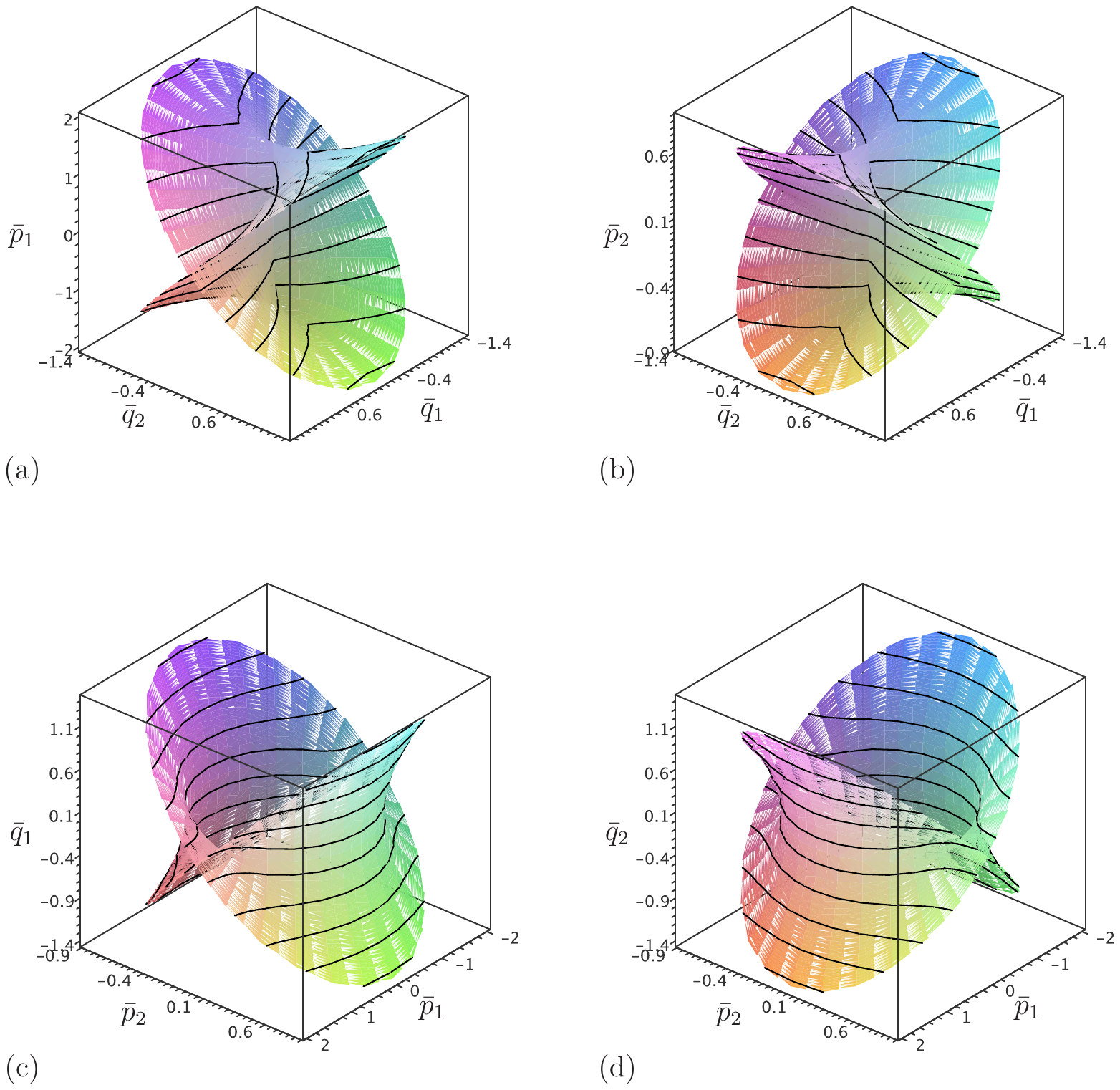}

 \vspace*{-6.5cm}
  FIGURE 2

\newpage

\hskip -3.0cm
\includegraphics[width=20.0cm]{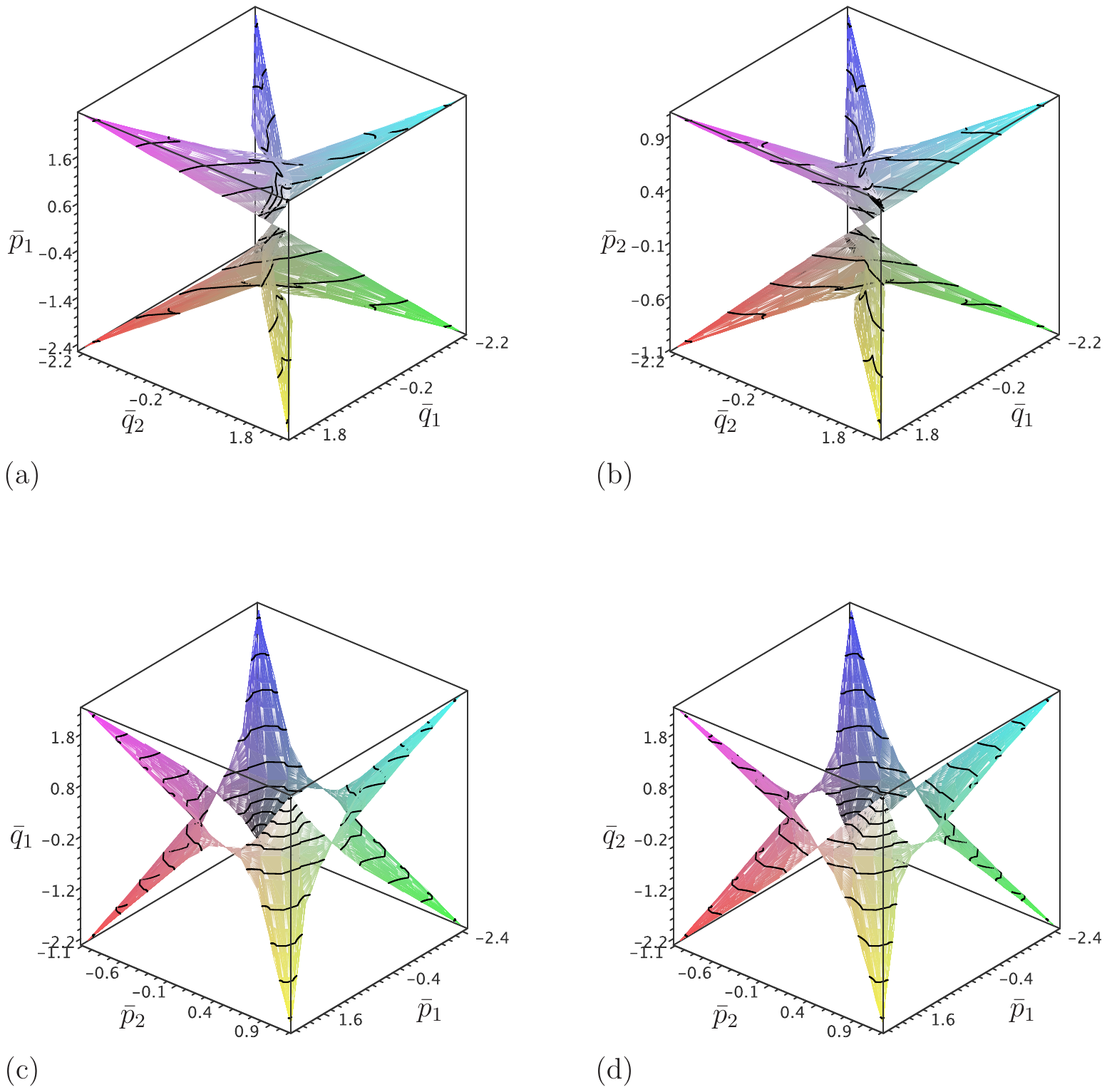}

 \vspace*{-6.5cm}
  FIGURE 3

\newpage

\hskip -3.0cm
\includegraphics[width=20.0cm]{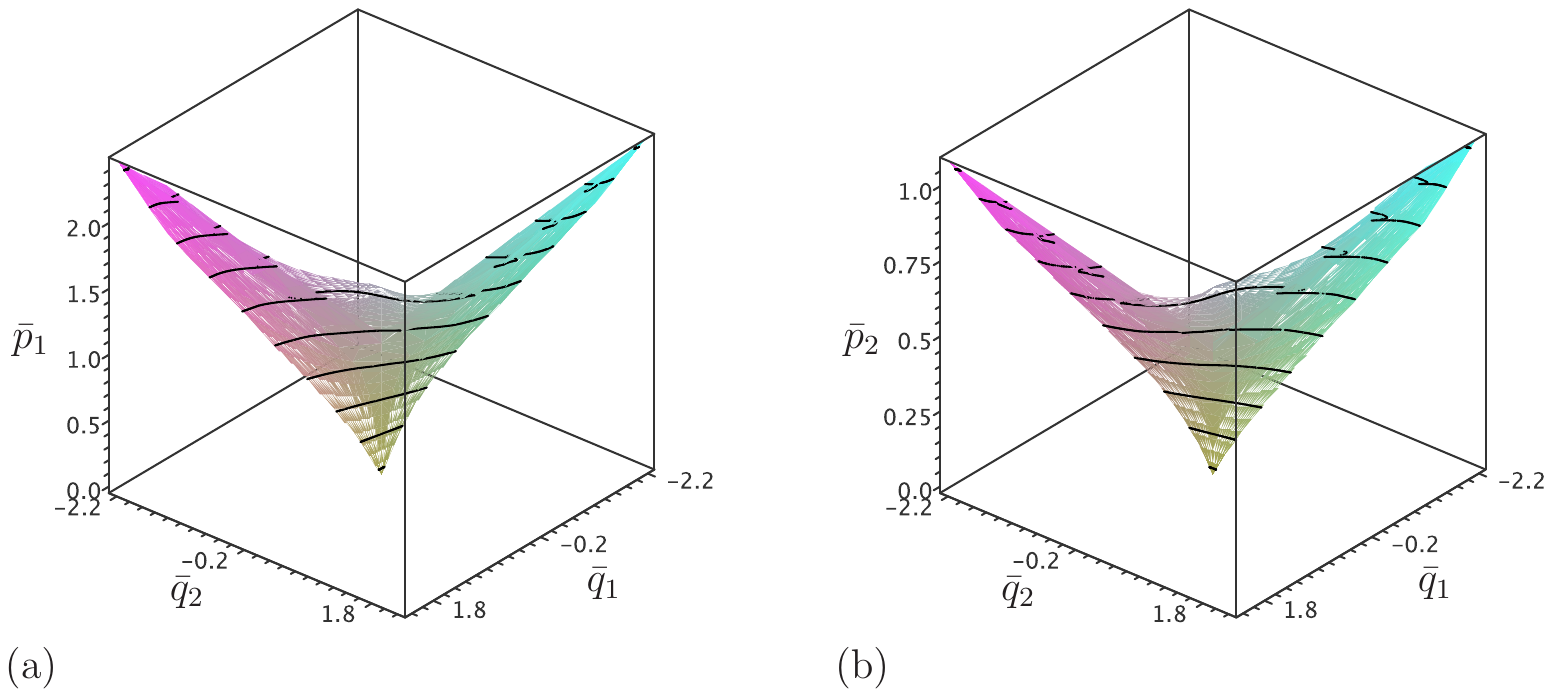}

 \vspace*{-6.5cm}
  FIGURE 4

\newpage

\hskip -3.0cm
\includegraphics[width=20.0cm]{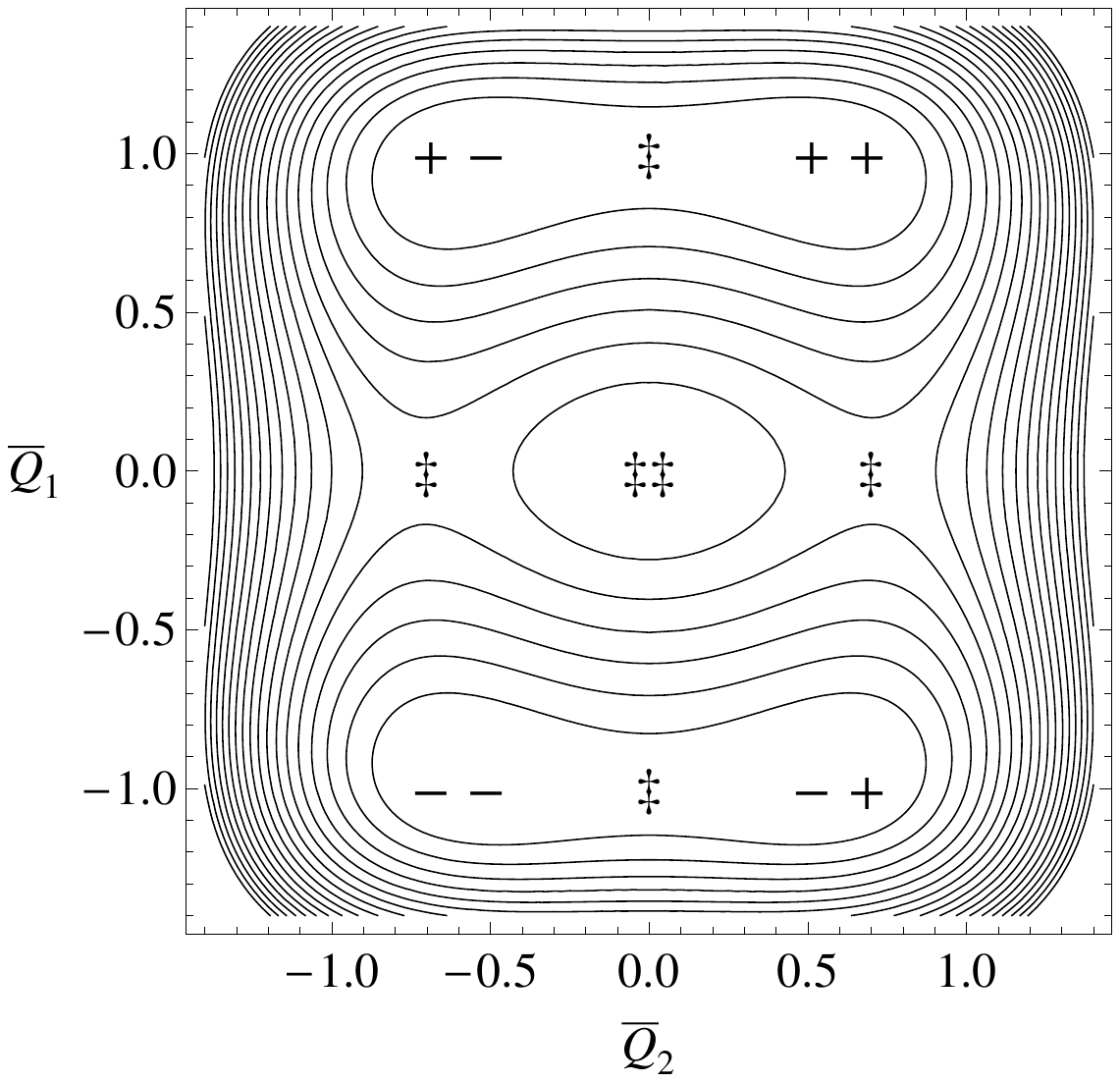}
  
   \vspace*{-6.5cm}
    FIGURE 5



  \newpage

\hskip -3.0cm
\includegraphics[width=20.0cm]{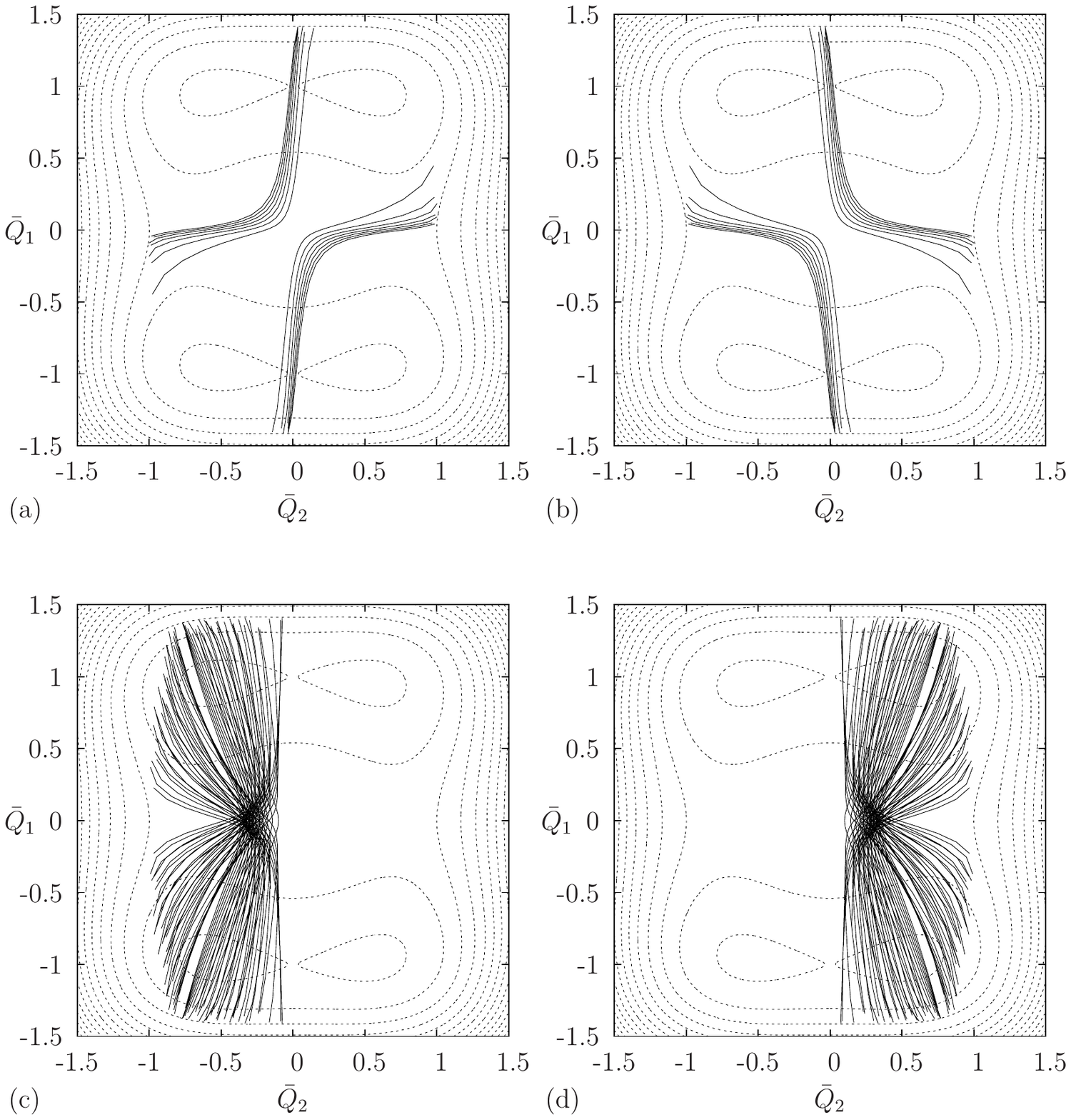}
  
   \vspace*{-6.5cm}
    FIGURE 6

   \newpage

\hskip -3.0cm
\includegraphics[width=20.0cm]{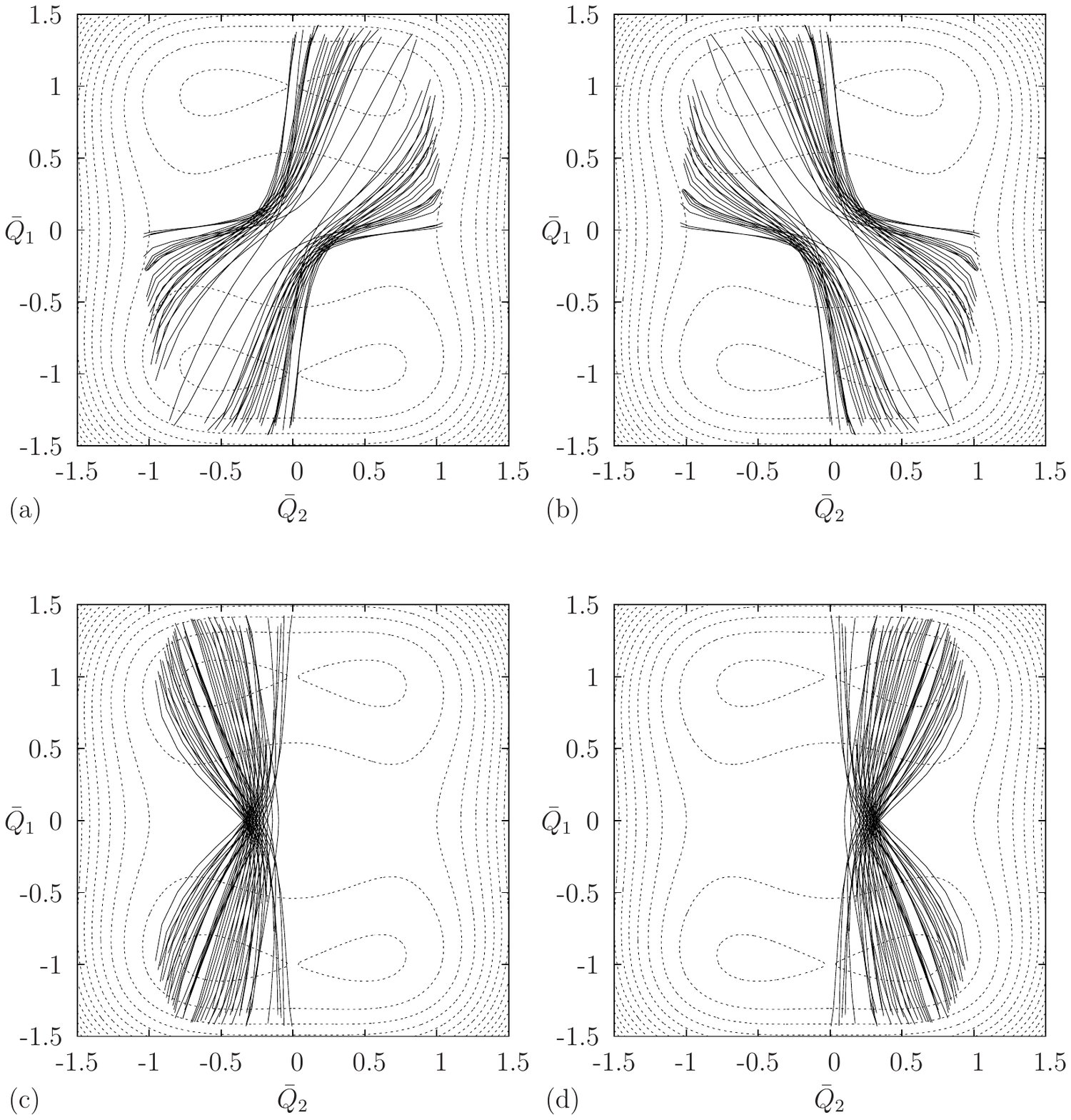}
   
     \vspace*{-6.5cm}
      FIGURE 7

  \newpage

\hskip -3.0cm
\includegraphics[width=20.0cm]{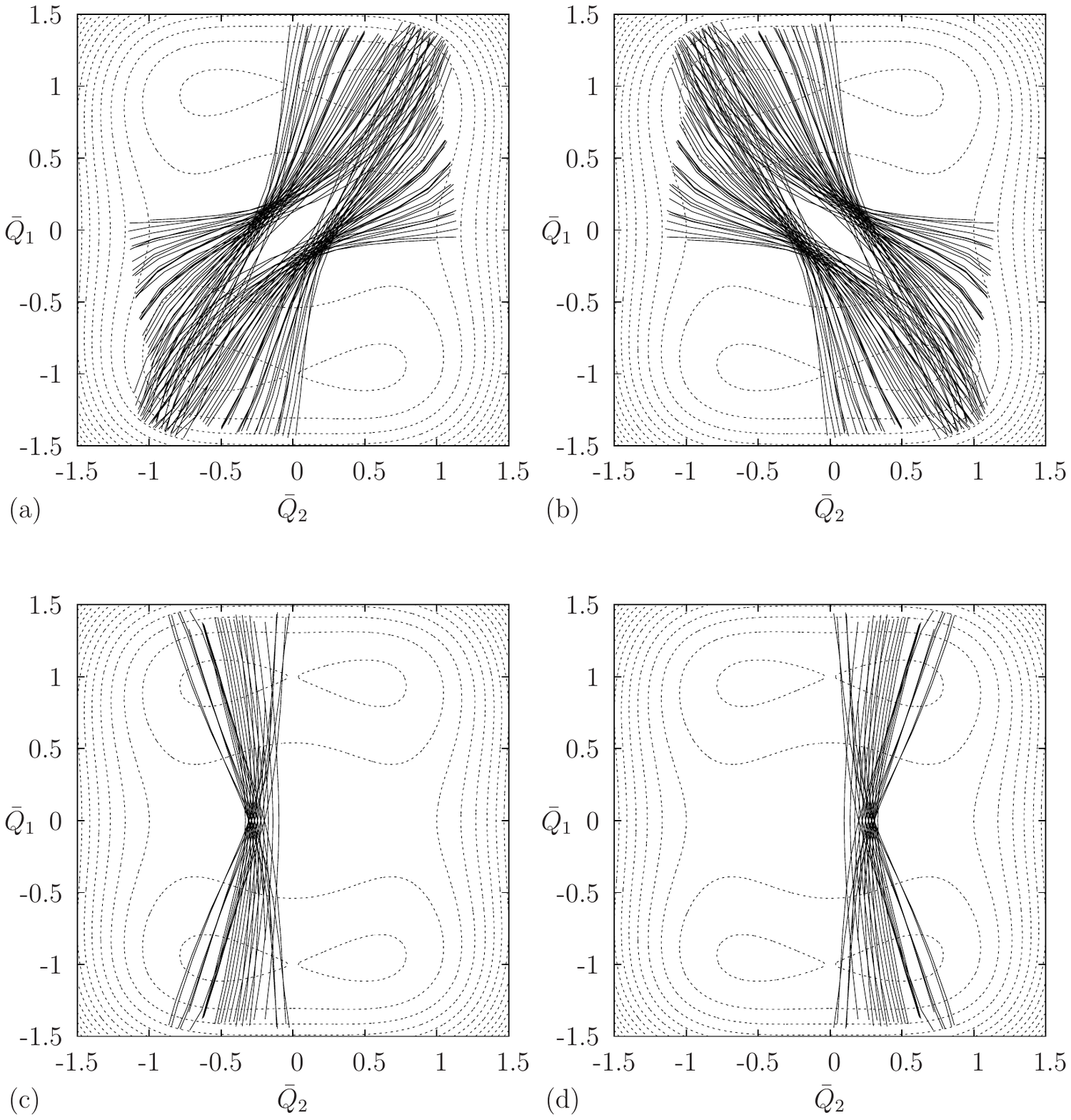}
 
     \vspace*{-6.5cm}
       FIGURE 8

 
 
 \newpage
 
\hskip -3.0cm
\includegraphics[width=20.0cm]{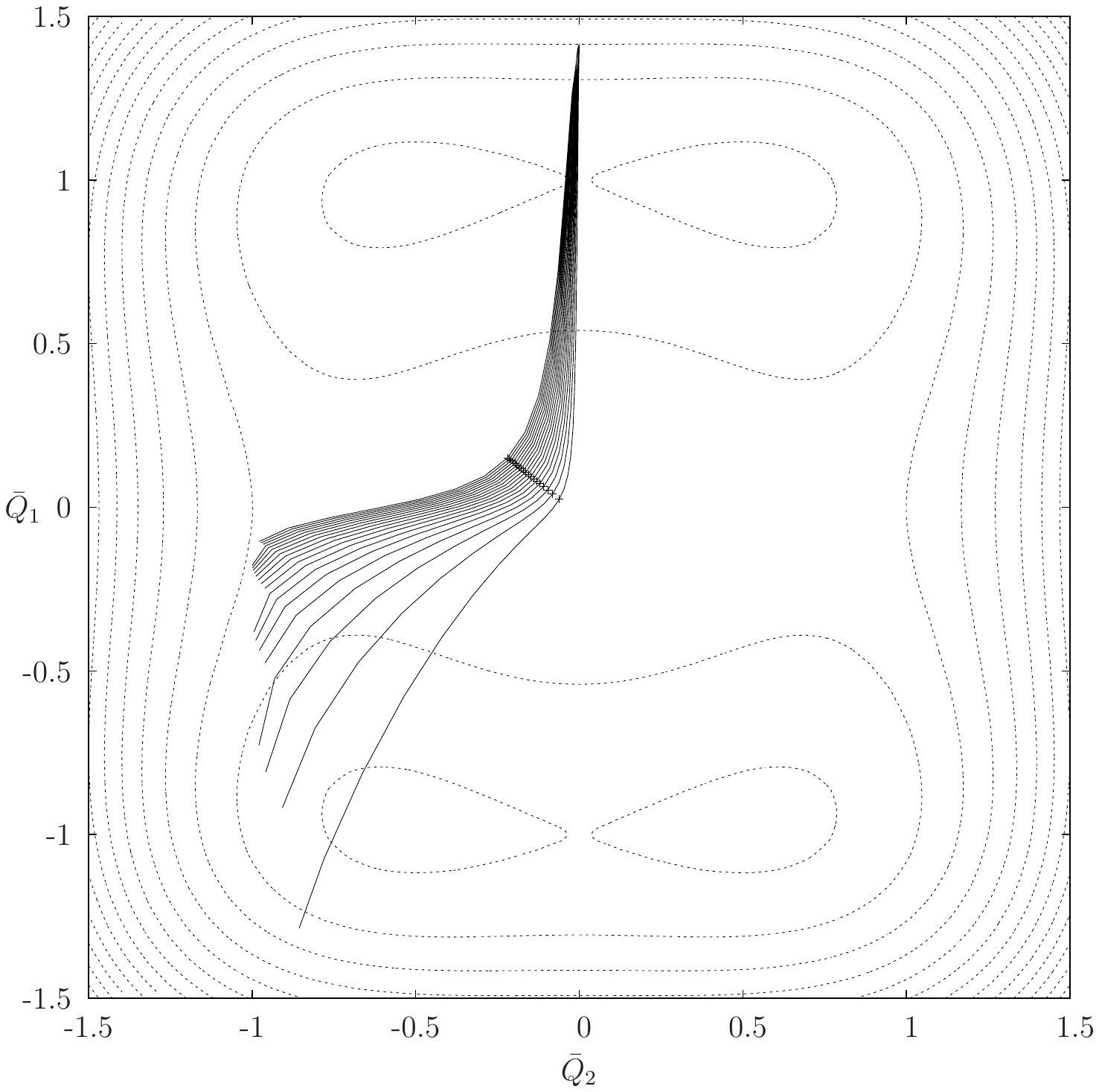}

\vspace*{-6.5cm}
       FIGURE 9


\newpage

\hskip -3.0cm
\includegraphics[width=20.0cm]{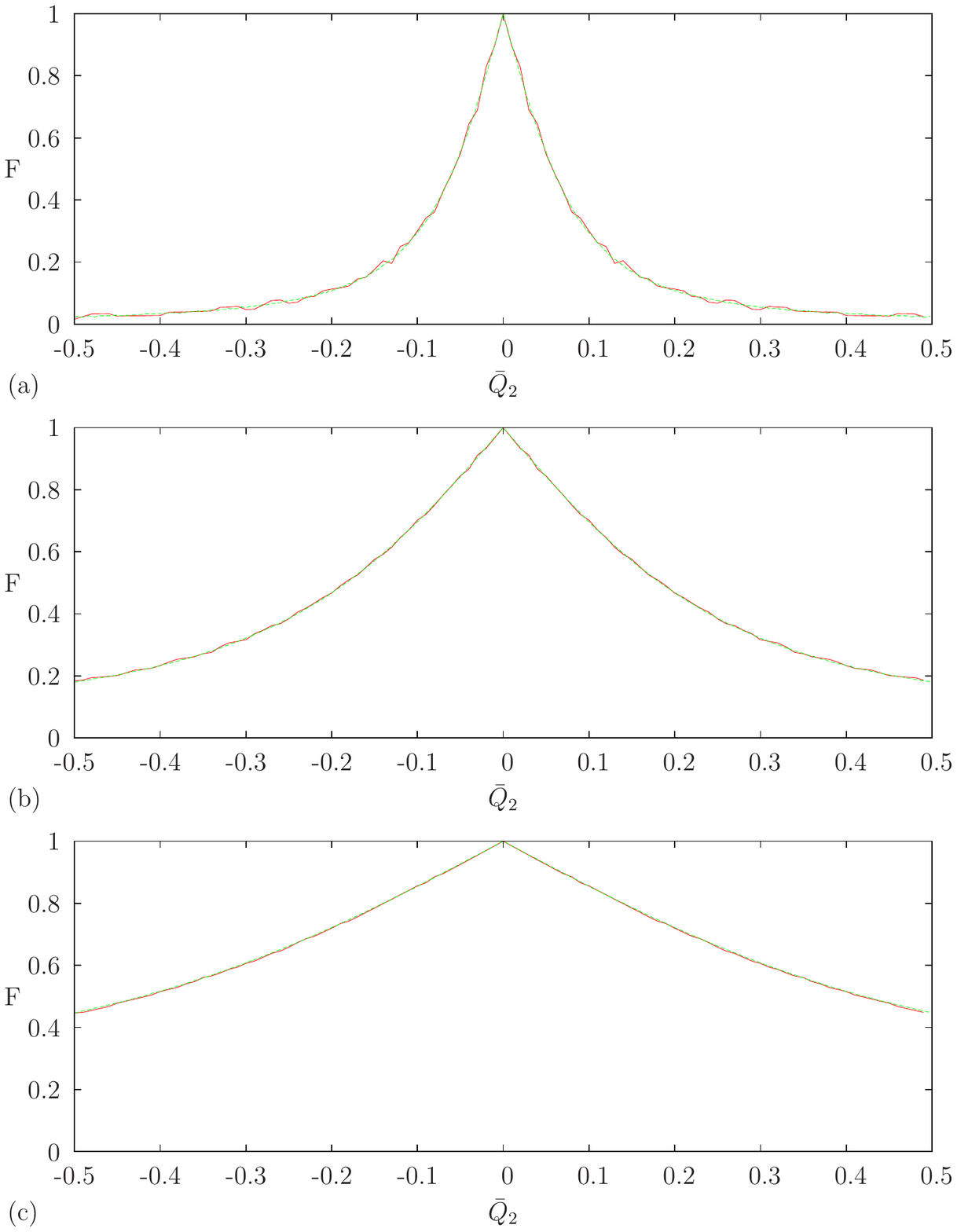}

\vspace*{-6.5cm}
       FIGURE 10



\newpage

\hskip -3.0cm
\includegraphics[width=20.0cm]{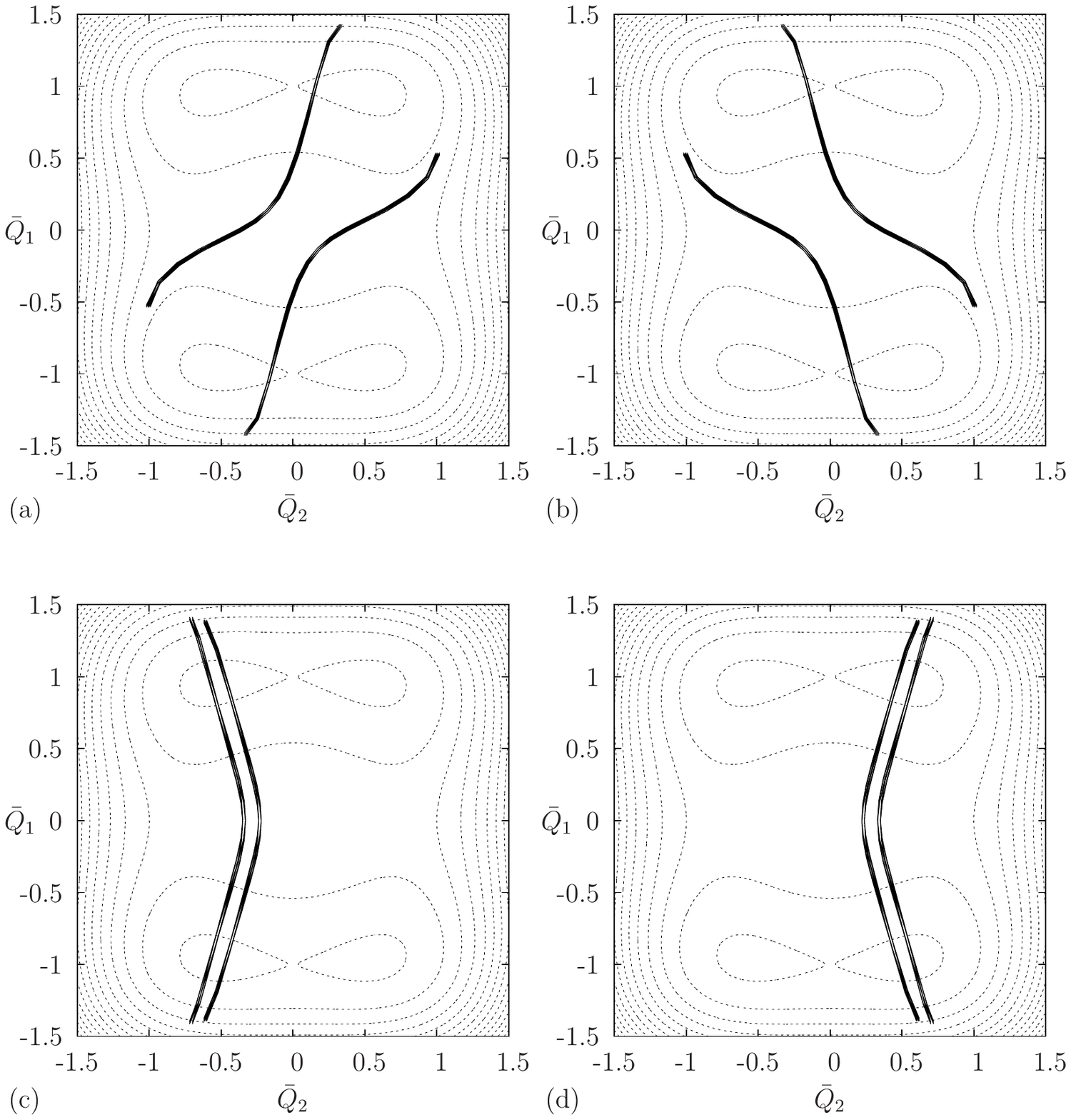}
 
 \vspace*{-6.5cm}
        FIGURE 11

 \newpage

\hskip -3.0cm
\includegraphics[width=20.0cm]{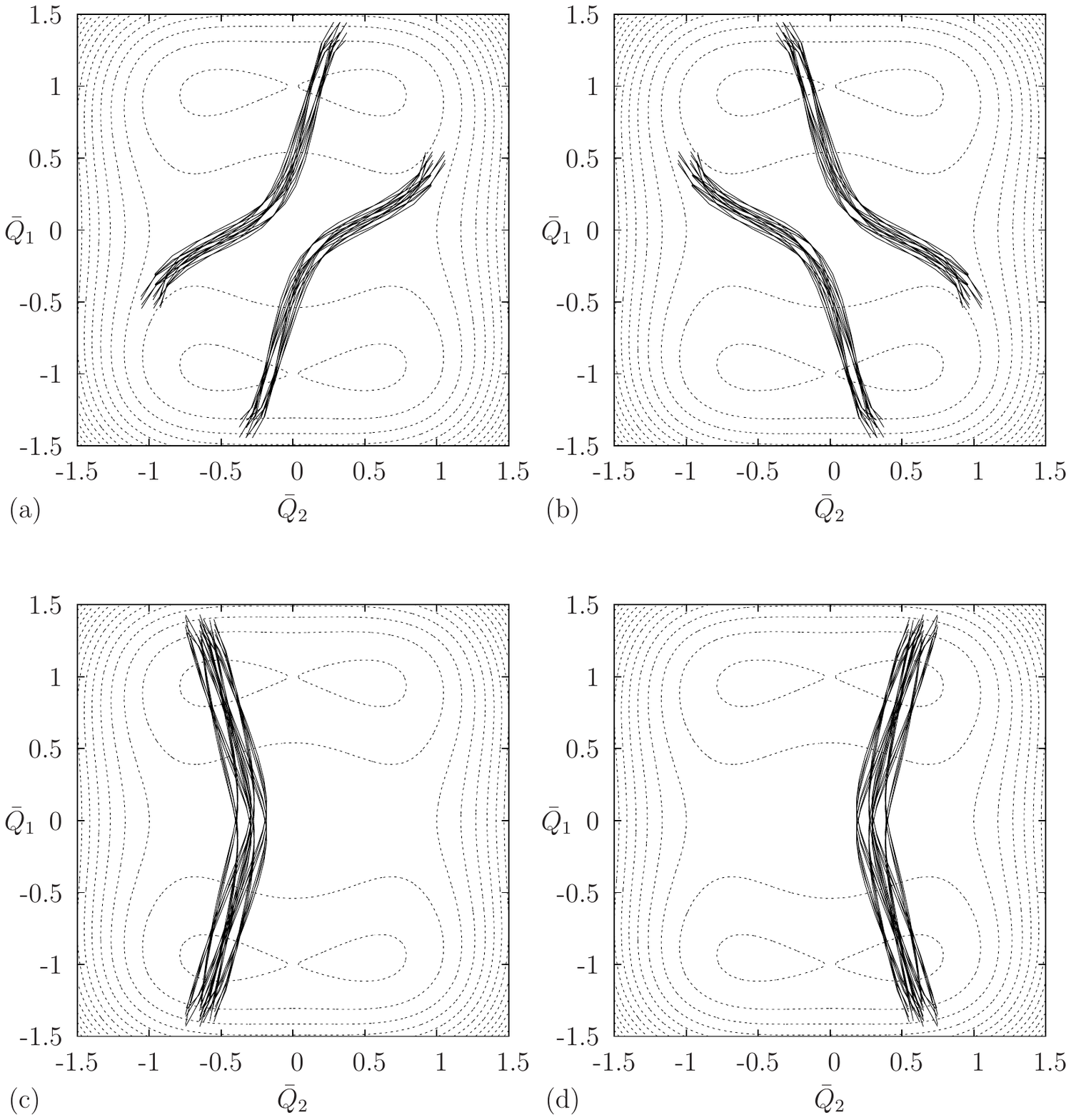}

 \vspace*{-6.5cm}
        FIGURE 12

\end{document}